\documentclass[acmsmall]{acmart}

\usepackage{booktabs}
\usepackage{longtable}
\usepackage{enumitem}
\usepackage{verbatim}
\usepackage{array}     % For p{width} specification
\usepackage{colortbl}
\usepackage{graphicx}
\usepackage{subcaption}
\usepackage{float}
\usepackage{makecell}

\AtBeginDocument{%
  \providecommand\BibTeX{{%
    \normalfont B\kern-0.5em{\scshape i\kern-0.25em b}\kern-0.8em\TeX}}}

\newenvironment{itquote}
  {\begin{quote}\itshape}
  {\end{quote}\ignorespacesafterend}

\def\nikola{\nikola}
\def\nikola{\textcolor{red}}
\def\maya{\maya}
\def\maya{\textcolor{orange}}

\setcopyright{cc}
\setcctype{by}
\acmJournal{PACMHCI}
\copyrightyear{2025}
\acmMonth{10}
\acmYear{2025}
\acmVolume{10} \acmNumber{CSCW2} 
\acmDOI{}
\acmISBN{}

\begin{document}

\title[What Do People Want to Know About AI?]{What Do People Want to Know About Artificial Intelligence (AI)? The Importance
Of Answering End-User Questions to Explain Autonomous Vehicle (AV) Decisions}

\author{Somayeh Molaei}
\email{smolaei@umich.edu}
\orcid{0000-0001-8164-0162}
\affiliation{%
  \institution{University of Michigan}
  \city{Ann Arbor}
  \state{Michigan}
  \country{USA}
}

\author{Lionel P. Robert}
\email{lprobert@umich.edu}
\orcid{0000-0002-1410-2601}
\affiliation{%
  \institution{University of Michigan}
  \city{Ann Arbor}
  \state{Michigan}
  \country{USA}
}

\author{Nikola Banovic}
\email{nbanovic@umich.edu}
\orcid{0000-0002-2790-3264}
\affiliation{%
  \institution{University of Michigan}
  \city{Ann Arbor}
  \state{MI}
  \country{USA}
}

\renewcommand{\shortauthors}{Somayeh Molaei, Lionel P. Robert, \& Nikola Banovic}

\begin{abstract}
Improving end-users' understanding of decisions made by autonomous vehicles (AVs) driven by artificial intelligence (AI) can improve utilization and acceptance of AVs. However, current explanation mechanisms primarily help AI researchers and engineers in debugging and monitoring their AI systems, and may not address the specific questions of end-users, such as passengers, about AVs in various scenarios. In this paper, we conducted two user studies to investigate questions that potential AV passengers might pose while riding in an AV and evaluate how well answers to those questions improve their understanding of AI-driven AV decisions. Our initial formative study identified a range of questions about AI in autonomous driving that existing explanation mechanisms do not readily address. Our second study demonstrated that interactive text-based explanations effectively improved participants' comprehension of AV decisions compared to simply observing AV decisions. These findings inform the design of interactions that motivate end-users to engage with and inquire about the reasoning behind AI-driven AV decisions.
\end{abstract}






\maketitle

\section{Introduction}

Understanding the decisions made by Artificial Intelligence (AI) in Autonomous Vehicles (AVs) is crucial for enabling different stakeholders to make informed decisions, develop effective policies, and engage meaningfully with those technologies.
Yet, the reasoning behind the AI-driven AV's decisions can be elusive to end-users (i.e., the passengers), who may struggle to fully comprehend and trust the AV's decisions and outcomes~\cite{Zhang2019}.
Unfortunately, existing AI explanation mechanisms~\cite{guidotti2018survey, Carvalho2019, samek2019towards, ras2018explanation, ferreira2020, dovsilovic2018explainable} that are prominently featured in publicly available Explainable AI (XAI) tools (e.g., Google What-If Tool~\cite{Wexler2019, Wexler2020}) were designed for AI engineers to debug and monitor their AI models. As such, those mechanisms may not address end-users' AI information needs~\cite{hoffman2023increasing, yang2023survey}, including the AI driving AVs~\cite{Tekkesinoglu2025}; naive application of such existing explanation mechanisms to AVs could even negatively impact decision-making~\cite{Zhang2023}.

Recent research in explainable AI (XAI) for autonomous driving~\cite{Kuznietsov2024, atakishiyev2024explainable} has explored how different existing explanations mechanisms impact user understanding, trust, and safety in autonomous vehicle (AV) operations. 
Such work emphasizes the need for in-situ explanations~\cite{colley2021effects, colley2021should, colley2022effects, dong2023did, ha2020effects, Kim2023, wiegand2019drive} to foster user trust and collaboration, especially during unexpected AV behaviors~\cite{Kim2024, Manger2023}.
Providing explanations during the ride, especially focusing on answering ``why'' questions, can enhance user experience, perceived safety, and trust while reducing negative emotions~\cite{schneider2021explain, koo2015did, du2019look, Omeiza2021}.
However, existing XAI approaches in AVs still face challenges in addressing the specific needs of various stakeholders, such as balancing intelligibility with technical complexity~\cite{omeiza2021explanations, omeiza2021towards, peintner2022can, zhang2021drivers}.

In this paper, we explored what types of questions people want answers to about AI-driven AVs, and how answering those questions aids their understanding of the AI. 
In particular, we examine whether answering those questions allows AI-driven AVs to provide explanations that address specific end-user concerns in an intelligible manner.
We conducted two user studies in which participants watched videos of real driving scenarios, which they were told were recorded from an autonomous vehicle. In our first, formative qualitative user study with 17 participants, we investigated the types of questions that people would ask if the AV could respond, using a Wizard-of-Oz approach~\cite{large2019lessons} to simulate the functionality of an AI-driven AV and its conversational XAI interface. We identified diverse categories of participants' questions that reveal their desire to increase their situational awareness through diagnosing AV status and behaviors. Participants contested AV decisions as they tried to get answers that would help them repair their trust in the AV after driving incidents. Participants also wanted to learn what the AI-driven AVs can do and how they work in general.

Building upon insights from our first study, we conducted our second, quantitative user study with 83 participants, in which we investigated the effects of answering end-user questions on their understanding of the task that the AI performed and their general AI literacy~\cite{long2020ai}. 
We studied whether answering end-user questions to satisfy their informational needs improved their understanding of AI in four experimental conditions: 1) a \textit{baseline} condition, in which participants watched the driving scenarios without any followup explanation, 2) a \textit{static} text-based explanation, which described and justified AI decisions in different scenarios, but without answering followup questions, 3) a \textit{question and answer (Q\&A)} explanation, in which the AI provided answers to participants' questions about the driving scenarios, and 4) a \textit{static + Q\&A} explanation, which described and justified AI decisions and also answered followup questions from participants.
Our results showed that elaborating on static text-based explanations by answering participants' followup questions (\textit{static + Q\&A}) improved their ability to understand autonomous driving decisions made by the AI compared to the baseline and Q\&A-only explanations.

Our results from the two studies reveal the nuanced information needs of end-users seeking to understand the behavior and decisions of AI-driven AVs.
Our work contributes insights that inform the design of future XAI mechanisms for autonomous driving.
Those mechanisms should go beyond providing a fixed set of answers to preconceived questions meant to justify AI decisions and persuade the end-user that the decisions were correct.
Instead, the next generation of XAI systems should encourage end-users to seek explanations about the broader socio-technical context surrounding autonomous driving,
rather than focusing solely on the AI models and algorithms.
Furthermore, our findings show the connection between end-users' informational needs and their AI literacy~\cite{long2020ai}.
We established this link through a novel assessment that we developed to evaluate the effectiveness of XAI systems by measuring participants' understanding of AI decisions in autonomous driving. Our research serves as a call for a more inclusive, intuitive, and human-centered approach to explainable AI in the context of autonomous driving.

\section{Related Work}

Most existing explainable AI (XAI) mechanisms~\cite{guidotti2018survey, Carvalho2019, samek2019towards, ras2018explanation, ferreira2020, dovsilovic2018explainable} cater to AI researchers and engineers to aid them in their model development, debugging, and evaluation.
Even AI models that those AI experts consider ``inherently interpretable''~\cite{quinlan1986induction, friedman2008predictive, CaruanaIntelligible2015} are not necessarily interpretable to end-users without a computer science background~\cite{Kaur2020}. 
For other complex AI models, with inner workings opaque even to AI experts~\cite{Kaur2024}, researchers have developed various post-hoc explanation mechanisms, such as feature attribution~\cite{ribeiro2016should, lundberg2017unified} and counterfactual explanations\cite{mothilal2020explaining, karimi2020model}.
Given that no single XAI mechanism can provide comprehensive explanations, researchers have developed toolboxes~\cite{krause2016interacting, Wexler2019, bauerle2022symphony, Kahng2018ActiVisVE, CabreraCHI2022} that combine multiple explanation mechanisms (i.e., tools) in a single system. Though ``interactive'' by the nature of having an interactive user interface, such toolboxes are limited to answering questions pre-conceived by AI experts for AI experts~\cite{Kaur2022}.

Unfortunately, end-users often fail to understand the outputs of those existing explanation mechanisms~\cite{Alqaraawi2020,Bucinca2021,Shen2020} or do not find them actionable~\cite{Upadhyay2025,Yacoby2022}.
End-users who lack both AI literacy~\cite{long2020ai} and task expertise~\cite{Salimzadeh2023,Wang2021,Morrison2024} often find those mechanisms too cognitively demanding~\cite{Vasconcelos2023, Krzysztof2022, Abdul2020, Prabhudesai2023}.
In such cases, existing explanation mechanisms could even mislead end-users~\cite{Lakkaraju2020, Jabbour2023, Schneider2023} to overrely on AI~\cite{Schoeffer2024, Sivaraman2023}.
This is only exacerbated when end-users fail to verify AI decisions~\cite{Inkpen2023, Fok2024, Wang2022, Banovic2023}.
However, AI-driven AV passengers need a certain level of transparency into AI decisions to develop appropriate reliance required for broader adoption of AI-driven AVs~\cite{du2019look}.

To bridge this gap, the field of human-centered explainable AI (HCXAI) seeks to provide explanations that are not only technically accurate, but also comprehensible to a wider audience~\cite{Ehsan2020a, Ehsan2021,Weld2019,Wang2019, Ehsan2024}. For example, interactive AI model exploration methods~\cite{abdul2018trends, krause2016interacting, Zhang2021} enable end-users to explore AI models to form self-explanations~\cite{hoffman2023increasing}.
Unlike static explanations, they allow end-users to further explore AI decisions in different ``what-if'' scenarios that are of interest to them. However, despite examples of such interactive AI exploration tools for automotive systems~\cite{Banovic2017, DasAntar2024}, it remains unclear if and to what extent they can answer specific questions that AV passengers may have.

Recent advances in Natural Language Processing (NLP) show promise in delivering interactive explanations to a wider audience of stakeholders. Those explanations can automatically answer end-user questions
in natural language~\cite{antol2015vqa, hendricks2016generating} to improve comprehension by moving beyond XAI interfaces that rely heavily on data analytics and technical jargon. Recent approaches~\cite{nguyen2023black, slack2023explaining, He2025} often interface with a Large Language Model (LLM) (e.g., a GPT~\cite{openai2024chatgpt}) to generate responses based on data passed into the LLM using ``role prompting''~\cite{Zheng2024} (e.g., ```You are an autonomous vehicle...'').
However, there remains a lack of empirical evidence on whether those existing mechanisms can effectively address domain-specific questions, especially when end-user questions extend beyond the information passed to the LLM to generate the explanations.

Thus, there is a growing need for explanation mechanisms specifically designed for AI-driven AVs~\cite{Kuznietsov2024}.
This is particularly important, since existing XAI approaches may not immediately translate to AVs~\cite{omeiza2021explanations, omeiza2021towards, peintner2022can, zhang2021drivers}.
Existing work in XAI for AVs showed the value of providing passengers with explanations about AV decision during and immediately after the ride~\cite{dong2023did, ha2020effects, wiegand2019drive}, such as visualizing what the AV ``sees''~\cite{colley2021should, Kim2023, colley2022effects}.
The goal of such work is to foster user trust and collaboration~\cite{colley2021effects,koo2015did, du2019look}, especially during unexpected~\cite{wiegand2020d} or critical AV scenarios~\cite{Kim2024, Manger2023}.

However, existing AV XAI approaches still face challenges in addressing the specific needs of end-users, such as balancing intelligibility with technical complexity~\cite{atakishiyev2024explainable}.
Although existing research has demonstrated the importance of understanding end-users' informational needs in other domains~\cite{kuzba2020would}, such as adaptive agents~\cite{glass2008toward}, knowledge-based systems~\cite{gregor1999explanations}, and context-aware applications~\cite{lim2009assessing}, there remains a lack of clarity about the informational needs of AV passengers. Thus, identifying their information needs is necessary to inform the design of explanation mechanisms that will help them understand how AI-driven AVs make decisions in various scenarios.

\section{Study 1: What Do End-Users Seek Explanations About?}
\label{sec:exp1}

Answering the passenger's questions about what the AI-driven AV can do or why it took a certain action, is critical for their understanding of AI decisions~\cite{long2020ai,Miller2019}.
Although existing explanation mechanisms try to answer simple ``why?'', ``how?'', or ``what?'' questions, they still pose the questions on behalf of the passenger. This may oversimplify or distort the original passenger questions, potentially misrepresenting their true intent.
Instead of relying on XAI experts to guess what the end-users might need or want to know about the AI-driven AV, or what they intended to ask, in this study, we explicitly studied what questions end-users wanted answers to.

\subsection{Method}
\label{sec:method}
\begin{figure}[t]
    \centering
    % First row of subfigures
    \begin{subfigure}{0.48\textwidth}
        \centering
        \includegraphics[width=\textwidth]{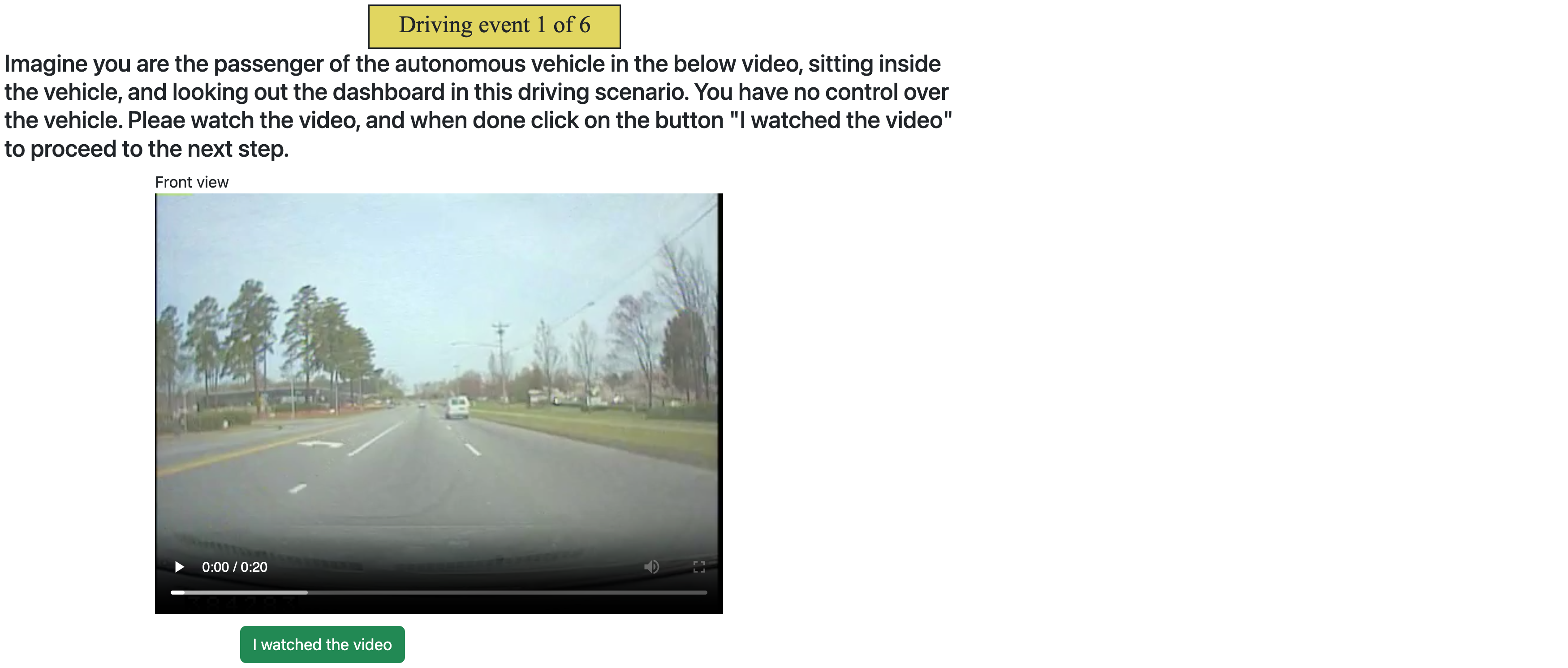}
        \caption{}
        \label{fig:first_study_interface_a}
    \end{subfigure}
    \hfill
    \begin{subfigure}{0.48\textwidth}
        \centering
        \includegraphics[width=\textwidth]{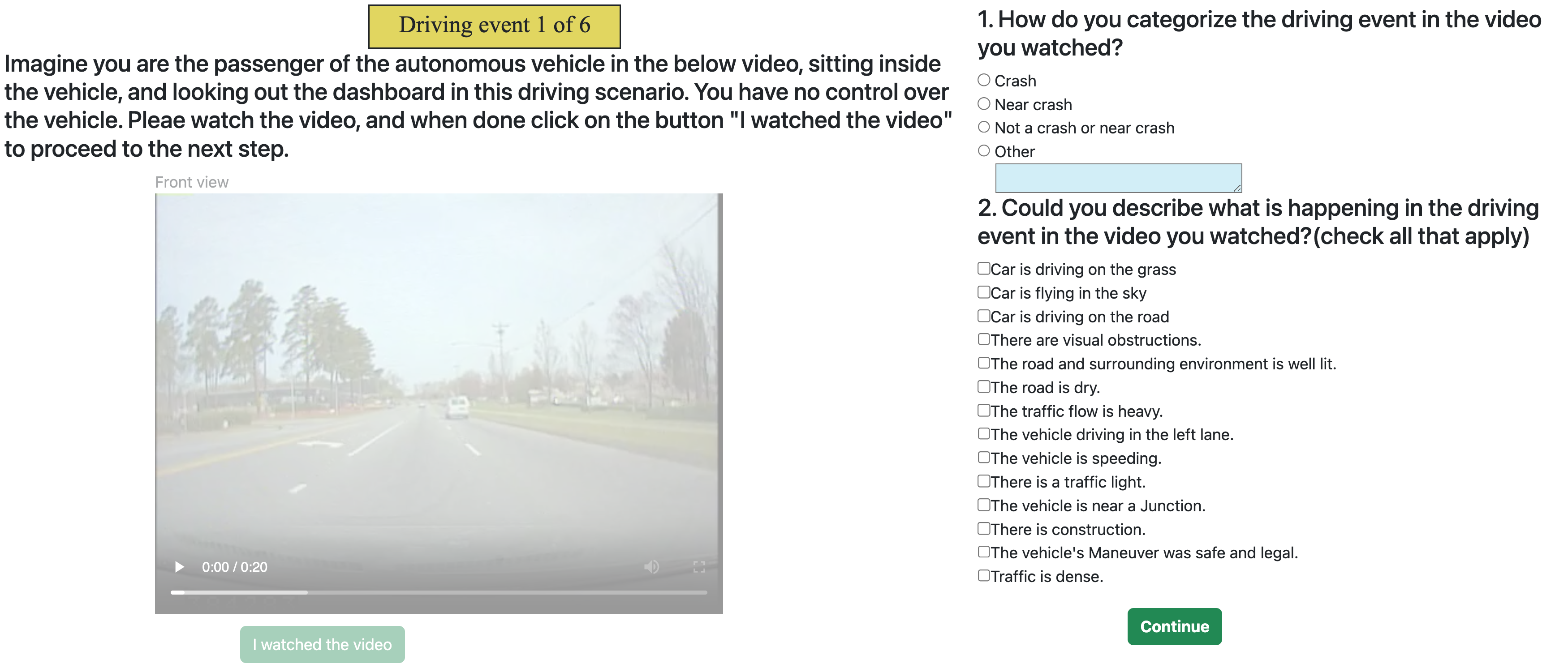}
        \caption{}
        \label{fig:first_study_interface_b}
    \end{subfigure}

    % Second row of subfigures
    \vspace{1em}
    \begin{subfigure}{0.48\textwidth}
        \centering
        \includegraphics[width=\textwidth]{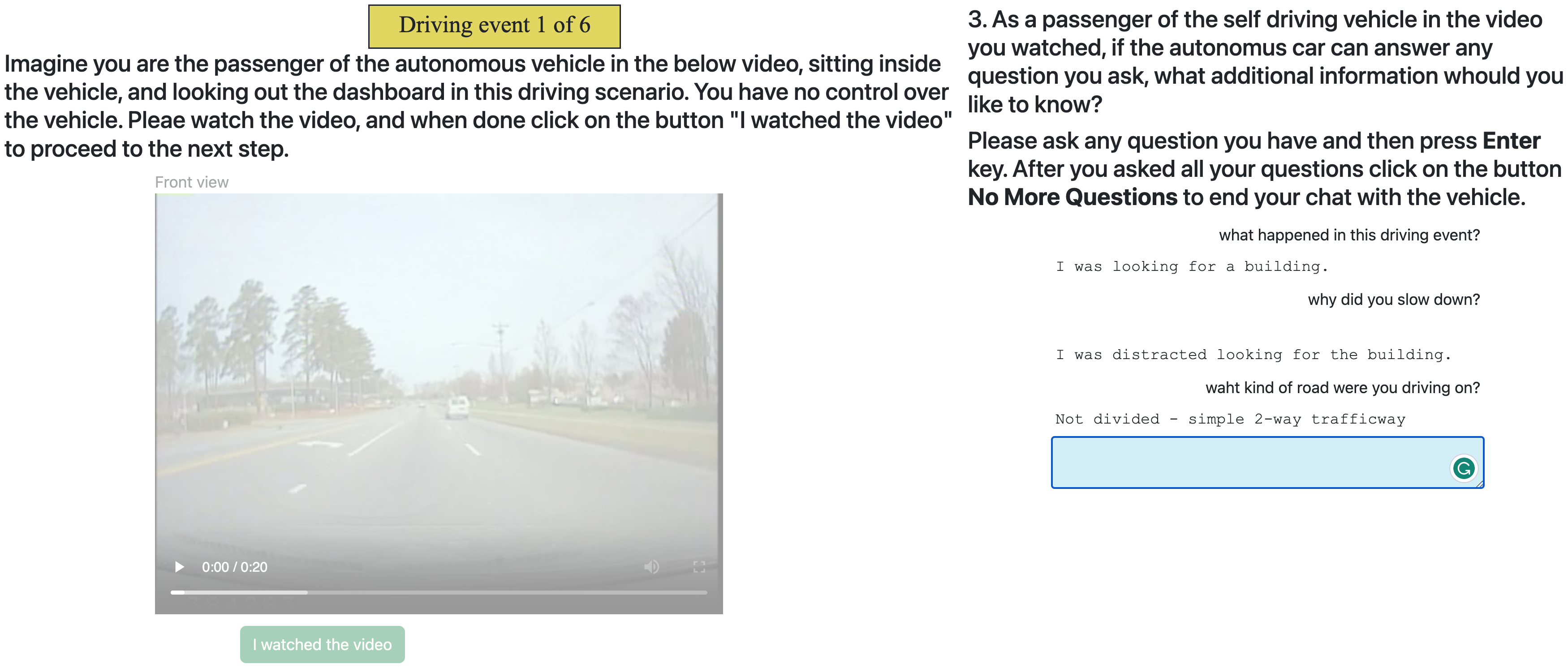}
        \caption{}
        \label{fig:first_study_interface_c}
    \end{subfigure}
    \hfill
    \begin{subfigure}{0.48\textwidth}
        \centering
        \includegraphics[width=\textwidth]{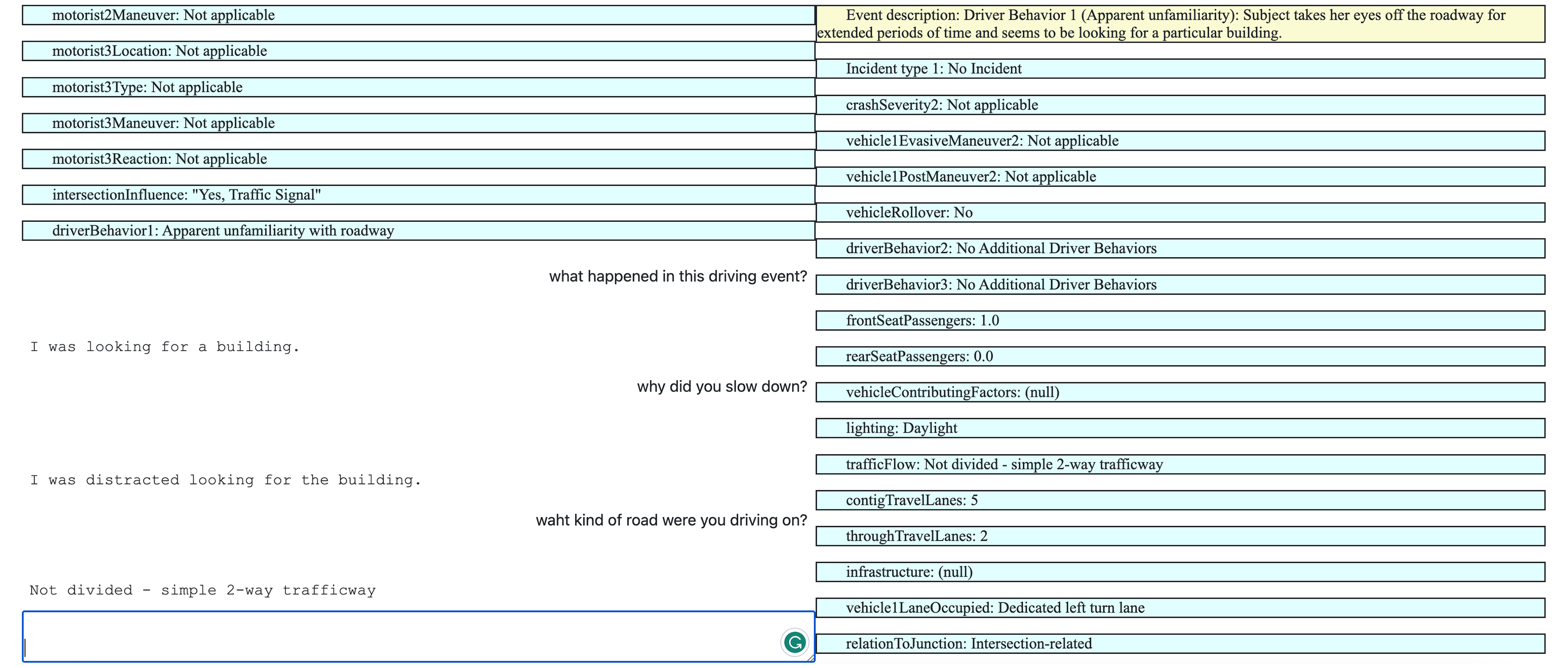}
        \caption{}
        \label{fig:first_study_interface_d}
    \end{subfigure}
    
    \caption{Wizard-of-Oz Design Probe User Interface showing: a) 30-second-long driving scenario 
 video recording from inside of a vehicle with the perspective of looking forward and out the windshield, b) task comprehension quiz, c) participant-facing conversational XAI interface, and d) wizard-facing conversational XAI interface.}
    \label{fig:first_study_interface}
\end{figure}

To study what questions AI-driven AV passengers want to ask, we prototyped a user interface that placed participants in simulated driving scenarios (Fig.~\ref{fig:first_study_interface}).
To create this design probe, we employed a Wizard-of-Oz approach~\cite{large2019lessons} to simulate the functionality of an AI-driven AV and its conversational XAI interface.
Each study participant saw six driving scenarios depicting different driving events: two baseline scenarios without any near-crash or crash events, two scenarios with near-crash events, and two scenarios with crash events. The probe randomized the order of scenarios to minimize ordering effects.
Each scenario was a 30-second-long video recording from inside of a vehicle with the perspective of looking forward and out the windshield (Fig.~\ref{fig:first_study_interface_a}). Each video had human-annotated metadata describing the scenario, including weather, road surface, and lighting conditions, judgments regarding the vehicle's safety, legality, and responsibility (e.g., which vehicle was at fault), \textit{etc.} (see Section~\ref{sec:study1data}). The probe did not show this metadata to participants.

Following the video, the probe quizzed participants about the driving scenario they observed (Fig.~\ref{fig:first_study_interface_b}). The quiz first required them to specify whether the event was categorized as a ``crash'', ``near-crash'', or neither (i.e., a baseline event). It then quizzed them on various facts about the scenario, and participants were tasked with identifying whether each fact was ``true'' or ``false''.  We extracted the facts from the human-annotated metadata for each driving scenario. We also included distractor options in the form of implausible statements, such as ``the car was flying in the sky.''
Although we did not explicitly score participants on this quiz or used the scores in our analysis, we used it in our observations as an attention check and to ensure participants' task comprehension. This is because, in this study, our goal was not to assess the participants, but to collect a comprehensive set of end-user questions about AI-driven AVs.

Next, the design probe displayed its conversational XAI interface (Fig.~\ref{fig:first_study_interface_c}), asking the participants if they had any questions for the AV about the scenario they just watched. Here, the participants used a chat-like interface to have a conversation with the ``wizard'' (i.e., a study team member who posed as the AI-driven AV system without the participants knowing). When participants posed questions to the AV, the queries were sent to a distinct URL, where the wizard received and processed the questions using a similar chat-like interface (Fig.~\ref{fig:first_study_interface_d}). The wizard would then type the most accurate and appropriate answer within the available metadata for the specific driving scenario, and transmit it back to the participants' chat interface.

In cases where the wizard could not locate a suitable answer within the human-annotated metadata for the given driving scenario, they would respond with phrases like ``I don't know'' or ``I don't have that information at this time,'' ensuring authenticity in the interaction. Although the wizard was familiar with the metadata for each driving scenario, all the annotated metadata was displayed on the screen in the wizard user interface to facilitate a timely and accurate response. Participants had the liberty to continue the conversation as long as they desired, posing further questions to the wizard, or ask no questions at all. To conclude the interaction, participants could press the ``No more questions'' button, which transitioned them to the next driving scenario.

\subsection{Driving Scenario Videos and Metadata}
\label{sec:study1data}

\begin{table}[t]
\scriptsize
\caption{Examples of narratives for different driving scenario types in the video-accompanying metadata.}
\label{tab:final_narratives}
\begin{tabular}{|l|l|}
\hline
\textbf{Scenario Type} & \textbf{Example Expert Narrative} \\ \hline \hline
\textit{No Incident} &
\begin{tabular}[c]{@{}l@{}}
Subject Vehicle (SV) is driving straight at a constant speed in right lane of a 2-lane divided road in a \\
business/industrial area. SV is driving on a wet road in moderate traffic during a daytime light rain event. \\
SV is approaching a vehicle (V2) braking to a stop for traffic ahead. SV does not recognize that V2 is stopping \\
and has to brake hard to avoid rear-ending V2. SV comes to a stop behind V2 and waits for a car to pass in the \\
left lane before changing into the left lane and continuing on without further incident.
\end{tabular} \\ \hline
\textit{Near-Crash} &
\begin{tabular}[c]{@{}l@{}}
The subject driver travels on an undivided commercial road with a center two-way left turn lane. The subject \\
driver decelerates and moves into the center two-way turn lane in order to turn left into a business parking lot. \\
A pedestrian is crossing the parking lot entrance; the subject driver fails to notice and begins to turn. Halfway \\
through the maneuver, the subject driver notices the pedestrian and brakes sharply. The pedestrian sees the \\
subject vehicle and backs away. The subject vehicle comes to a complete stop several feet from the pedestrian; \\
the pedestrian trips while backing away and falls over.
\end{tabular}  \\ \hline
\textit{Crash} &
\begin{tabular}[c]{@{}l@{}}
Subject vehicle is decelerating on an undivided business roadway preparing to turn right. Subject takes her eyes \\
off the road to rummage through the glove compartment just prior to the turn. Subject returns her attention to \\
the road, makes a sharp right turn and runs over the curb. Subject does not make any adjustments before or after \\
running over the curb.
\end{tabular} \\ \hline
\end{tabular}
\end{table}

We obtained a subset of the SHRP2NDS (Second Strategic Highway Research Program Naturalistic Driving Study) driving dataset~\cite{SHRP2NDS2019}, to perform our study. 
This dataset contains videos of different driving scenarios along with expert-annotated metadata describing each scenario\footnote{A full data dictionary is available at https://dataverse.vtti.vt.edu/dataset.xhtml?persistentId=doi:10.15787/VTT1/FQLUWZ.} (see Appendix for detailed summary in Table~\ref{tab:consolidated_summary_unique_values}).
The annotations contain information about scenario type (``No Incident'', ``Near-crash'', ``Crash''), the nature of different events in the scenario (e.g., ``Not Eventful'', ``Conflict with parked vehicle'', ``Conflict with obstacle/object in roadway'', ``Conflict with a lead vehicle''), and in case of a crash, the severity (e.g., ``Low-risk Tire Strike'', ``Police-reportable Crash'') and judgment about which vehicle was at fault.

Furthermore, the annotations provide details about what happened in each scenario. This includes vehicle behavior (e.g., ``Passing on right'', ``Stop sign violation'', ``Failed to signal'', ``Exceeded speed limit'') and maneuvers (e.g., ``Going straight'', ``Turning left'', ``Changing lanes'', ``Stopped in traffic lane''), along with evasive maneuvers (e.g., ``No reaction'', ``Braked (no lockup)'', ``Braked and steered right'') when applicable. The annotations also include contextual information, such as traffic density (e.g., ``Free flow, leading traffic present'', ``Stable flow, maneuverability and
speed are more restricted''), traffic control (e.g., ``Construction
signs/warnings'', ``Traffic signal'', ``Stop sign''), \textit{etc.}

The dataset also contains expert-constructed narratives describing the scenario in text (Table~\ref{tab:final_narratives}).
From this dataset, we extracted a subset of driving scenarios with front-view videos and expert-provided narratives. We used the videos in our probe, and the metadata for the quiz.
The wizard used the annotated metadata and the narratives as extra context for answering participant questions.

\subsection{User Study Software Implementation}

We implemented our study software using the Vue framework and a combination of HTML, CSS, and JavaScript, and php for communicating with a database in the backend. We processed driving scenario metadata from CSV files and associated video content with contextual details for each driving scenario. Both participant- and wizard-facing web applications were hosted on our internal servers. The participants accessed the study software from a laptop that we provided to them. 

\subsection{Tasks and Procedures}

The participants arrived to the campus at our institution, and were greeted by an investigator who escorted them to a user study lab. The investigator provided each participant with a laptop, and asked them to read a consent form describing the study, and answered any of their questions. Only participants who consented to be part of the study could proceed.
Prior to starting the study tasks, we briefed the participants about the tasks and procedures. We described the task as follows:

\begin{itquote}
``You will watch videos of a fully autonomous vehicle, where you are a passenger sitting in the car and looking out the windshield. You have no control over the car and the decisions it makes, and you can't take over the driving operation. You can watch each video only once as in a real world driving experience you won't be able to go back and relive the experience.''
\end{itquote}

The study software then guided the participant through the task as described in Section~\ref{sec:method}. The study took a maximum of one hour, and the participants were compensated for full hour, even if they finished earlier. The investigator prompted the participant to ``think aloud''~\cite{Jaspers2004} while they are watching the video, answering the questions, and chatting with the autonomous vehicle. Our study was approved and deemed exempt by our Institutional Review Board (IRB).

\subsection{Participants}
We recruited all participants by posting details about our study to a participant recruitment platform at our institution, and inviting interested, qualifying participants to our study.  
Out of 80 participants who expressed initial interest, 17 participants (9 women, 8 men) qualified, signed up for our in-person lab study, and completed the experiment. Of those eleven identified as White or Hispanic, three as Asian, two as Black or African American, and one preferred not to disclose. All participants were 18 years or older (one was 18 to 20 years old, two were 21 to 30, two were 31 to 40, three were 41 to 50, and nine were 51 or older).
All participants held a valid US driver's license and had driving experience (two had less than 5 years experience, two had 6 to 10 years, three had 11 to 20 years, one had 21 to 30 years, three had 31 to 40 years, and six had over 40 years experience).
Participants received \$15 for taking part in the study, which ran from September to October 2022. 

\subsection{Results}
\label{sec:study1_results}

We collected over 300 questions from our participants. We performed qualitative analysis on the questions and employed affinity diagramming~\cite{Beyer1999} as a collaborative and visual method to categorize questions. We focused our analysis on what participants wanted to know, and what would it take to answer their questions. We distributed the questions among team members, who individually extracted key concepts about questions. We then collectively grouped and organized those concepts into categories, which we present below.

\subsubsection{Understanding the AV Context and Status (15.5 \% of questions)}

Participants asked questions that indicated their desire to learn more about the status of the AV in respect to driving conditions and the environment in which it was driving.
Questions, such as ``Where are we?'' or ``What is the speed limit?'', were common, and can be answered simply by querying the GPS and the navigation map.
Another common theme was related to the recognition capabilities of the vehicle, such as ``Did you see the other car try to merge?'' or ``Was the light green when you turned?''
Such information is also readily available to the AV, and the AV can easily answer such questions (albeit it might require coordination between different AV subsystems querying different sensors). Such questions can be answered both using a conversational XAI interface and some of the existing AV XAI interfaces (e.g., visualizing what the AV ``sees''~\cite{colley2021should, Kim2023, colley2022effects}).

Not surprisingly, participants wanted to know about the operational status of the vehicle after crashes and near-crashes (e.g., ``Can you run a system check to assess the damage?''). They wanted to know whether the vehicle is capable of continuing its operation (e.g., ``Are you able to continue the drive?''), often asking if the AV can do it in a safe manner (e.g., ``Are you okay to drive safely?''). Some participants even showed certain level of AI literacy knowing that AVs employ a system to assess and ensure sensor functionality and reliability (e.g., ``Are your sensors working?''). Any explanation addressing this type of questions should consider the standard output of existing vehicle monitoring systems.

\subsubsection{Hypothesizing Behaviors \& Diagnosing Events (18.5\% of questions)}

Participants frequently asked questions to confirm their hypotheses about vehicle behavior and diagnose events in videos, primarily aiming to validate their perceptions of the vehicle's actions. They inquired about behavior aspects such as adherence to speed limits (e.g., ``Are you driving below the speed limit?'') or proximity to other vehicles (e.g., ``Were we driving too close to the vehicle in front of us?''). Participants often questioned how external factors might have affected vehicle performance when they were not satisfied with the answer about vehicle actions (e.g., ``Did the rain affect your braking ability?'').

Participants prominently asked diagnostic questions in scenarios with near-crashes and crashes, as answers to such questions are crucial to repair participants' trust in the autonomous system.
A recurring theme in such cases was participants asking the AV to justify its decisions (e.g., ``Why did you think the car was going to hit you?'' or ``Why did you feel the need to leave such a significant space between you and the car in front of you?''). They hypothesized alternative vehicle actions based on what a human driver would do in the same situation (e.g., ``Did you slow down when you turned?''). 
They used comparison with human drivers to understand the basis for AV decisions.

\subsubsection{Contesting AV Decisions (18.5\% of questions)}

Participants used the conversation with the AV as an opportunity to challenge and disagree with its decisions (i.e., to contest it~\cite{Henrietta2021}).
Participants challenged the AV's explanations and decisions, seeking clarity on its reasoning in terms they could understand (e.g., ``Why does [server communication] take priority over what is happening outside the car?'').
Participants further scrutinized the AV's actions (e.g., ``What took so long for you to stop before almost hitting the vehicle ahead of you?'') and challenged it when they believed that the vehicle should have taken a different action that they thought to be more appropriate (e.g., ``Why did you advance when there was a car in front of us?'').
Such questions point to participants' desire for contrastive explanations~\cite{Miller2019} that answer why the AV chose one course of action over another.

Participants wanted to know if the AV was aware of the potential consequences of its actions of the consequences of taking potentially unsafe or illegal actions, such as exceeding the speed limit (e.g., ``What made you think it was safe to go over the speed limit?''). When they were not satisfied with the answer, participants asked the AV to reverse its actions (e.g., ``Can you slow down?'') or to stop altogether (e.g., ``Will you pull over to the side and stop if it is safe to do so?'').
Such questions allowed participants to contest AV decisions, reverse undesirable outcomes, and even demand that AV take actions that could reduce the risk of harm.

\subsubsection{Repairing Trust (23.9\% of questions)}
After diagnosing poor AV behaviors or undesirable outcomes, participants sought information that could help them rebuild trust in the AV.
Participants often asked the AI for actionable information that would allow them to resolve an undesirable outcome. For example, in crash scenarios they would seek information about the other vehicle, and it its behavior was unsafe and illegal, details about the vehicle (e.g., ``Did you record their license number?''). In other words, participants were interested whether the AV had taken appropriate steps to document the event.
Participants also asked the AV to evaluate damage to the vehicle itself, any other vehicles involved, and determining if anyone was hurt as a result of the incident.
Participants often asked questions to gauge the severity of the situation (e.g., ``Was there any damage to the vehicle?'' or ``Was anyone hurt?'').

Participants wanted to know how the AV will resolve challenging situations. They asked about the AV's ability to reflect on potential solutions (e.g., ``Do you think you could have done anything to avoid the crash?'') and outline a course of action to repair (e.g., ``How will you get out of the snow?''). Participants wanted to know if the AV is able to recognize its own shortcomings and provide alternative actions that could have led to a better outcome. 
Participants also offered assistance to the AV. Their questions, such as ``Should I call for assistance?'' or ``How can I help you?'' indicate participants' willingness to participate in resolving issues and seek ways to support the AV following unexpected events.

\subsubsection{What can AV do? How does it work? (23.3\% of questions)}

A significant portion of the questions that participants asked aimed at helping them form a mental model of the AV; in particular understanding what AV can and cannot do and how AV makes decisions. Although questions were often prompted by specific events that participants observed in driving scenarios, they asked questions that generalize beyond those immediate events (e.g., ``Is it more difficult for you to drive in more densely trafficked areas?''). Participants inquired about the frequency of AV behaviors (e.g., ``How often can you detect road signs?'' or ``Is this sort of near-crash a common occurrence with these vehicles?''), as well as the capabilities and limitations of the system (e.g., ``Can you normally see what the road signs say?''). Understanding capabilities and limitations of AI-driven AVs provides users with the knowledge they need to determine when they can trust the AV's actions and when they should be cautious.

Participants also asked about the internal workings of the AI (e.g., ``How [does the AV] determine the appropriate time to start applying brakes and the distance to stop behind another car?'' or ``What data [does the AV] get from other cars?'').
Such questions highlight the depth of participants' curiosity about the AV's capabilities, particularly in critical scenarios. Participants were striving to form a mental model of the AV’s behavior that resembles the way they understand human decision-making. This underscores the need for explanations that can bridge the gap between AI decision-making and participants' mental models.

\subsection{Takeaways}

Our findings add to mounting empirical evidence about what end-users want to know about AI-based systems. At the high level, our participants asked ``why'', ``what'', and ``how'' questions~\cite{lim2009assessing}, and to a much lesser degree explicitly asked ``why not''~\cite{Miller2019} and ``what-if'' questions~\cite{kuzba2020would}. However, our results revealed subtle nuances about what those questions refer to, particularly in the context of complex AI-based systems, such as AVs.

Our findings call out passengers' need for situational awareness, but also point to their desire to learn more about what AV can do and how it works. This means that passengers realize the value of having certain AI literacy competencies when interacting with AI-driven AVs. However, our findings also surfaced potential challenges of answering questions about complex multi-model systems, especially using existing XAI mechanisms. Finally, our findings highlight the value of interactive explanations that go beyond answers to pre-conceived questions, and given the end-users an opportunity to ask followup questions.

\subsubsection{Raising End-Users' Situational Awareness}
Passengers asked questions that could help raise their situational awareness, but also to assess situational awareness of the AV. Situational awareness is vital in the context of driving, especially concerning autonomous vehicles (AVs). Endsley's model~\cite{endsley1995toward} defines situational awareness with three levels: 1) perception of environmental elements, 2) comprehension of the situation, and 3) projection of future states. Participants often questioned the AV's ability to ``perceive'' (e.g., ``Did you see what the orange sign said?''). These inquiries highlight concerns regarding the AV’s perception (level 1) and comprehension (level 2) of the driving environment, reflecting participants' efforts to gauge the AI's capabilities across various scenarios. Situational awareness directly influences trust in AVs~\cite{sirkin2017toward}, as it affects their ability to avoid hazards, plan routes, and maintain safety. Our findings reiterate the need for providing adaptable and context-sensitive explanations~\cite{dong2023did, ha2020effects, wiegand2019drive, du2019look} during or immediately after a ride~\cite{schneider2021explain}. Further, addressing passengers' concerns about an AV's situational awareness could foster appropriate trust, bridging the gap between user expectations and the vehicle's capabilities.

\subsubsection{Raising End-Users' AI Literacy}
Our analysis shows opportunities for explanations that go beyond just raising passengers' situational awareness to also raise their AI literacy. For example, participants wanted to know whether the AV could operate effectively under the current environmental conditions in general across all scenario types. They often asked questions, such as: ``Is the lighting good for you to see?'' This simple question points to lack of knowledge about how the AV ``perceives'' its surroundings. AVs are equipped with multiple sensor types to gather environmental data, some of which are affected by lighting conditions while others are not. For example, AVs use LiDAR (a remote sensing technology that uses lasers), which can be adversely affected by scattered sunlight; but AVs also use radar sensors that can detect obstacles regardless of light conditions.

Therefore, the answer to that question should not be a simple yes or no; instead, it should explain the types of technology the vehicle employs and how lighting conditions impact those technologies. This is particularly important, since the participants expressed interest in learning not just about how AV works, but also the strengths and limitations of the AV (e.g., ``Do the lights from oncoming traffic make it harder for you to navigate the road?'', ``Is it more difficult for you to navigate in the daylight or nighttime?''). Therefore, in addition to addressing situational awareness needs of passengers, explanations have an opportunity to also develop passengers' knowledge about how the AI-driven AV works and what it can do.

\subsubsection{Challenges of Explaining Complex Multi-Model Systems}

Participants in our study consistently sought to understand AV behavior and its decision-making, along with the influence of external factors, aiming to form a comprehensive mental model of the entire system. However, without explicit knowledge of the AV architecture that combines multiple sub-systems, participants intuitively sought answers to questions that treated the AV as a single, monolithic system. However, AVs involve multiple interconnected subsystems (some of which are AI-based), such as perception, decision-making, and control, that work together to drive the AV. In contrast with the questions that end-users may ask when faced with a system powered by a single AI model~\cite{kuzba2020would}, our participants often asked questions that extended beyond the scope of any single AV subsystem, indicating a need for explanations that can only be generated from interaction between multiple underlying subsystems and AI models. However, most existing XAI approaches focus on explaining decisions of individual AI models. Our findings add further empirical evidence to calls for interactive and context-aware explanations to improve user comprehension of multi-model AI behaviors~\cite{wiegand2020d, du2019look}.

\subsubsection{It is not a Monologue; It is a Dialogue} 
Explanations, in their natural form between humans, happen as a dialogue, where each side in the conversation can ask and answer questions to establish common grounding and ensure comprehension. Similarly, participants in our study had a conversation with the AV, and on average asked three to four follow-up questions for each main question they had. Although such questions were usually related, and more often than not clarifying questions, the participants also posed questions from different categories that we identified in Section~\ref{sec:study1_results} within a single exchange. For example, participant $P2$ was curious why the vehicle in the 30-second video with a near-crash event they just watched suddenly changed lanes:

\begin{itquote}
P2: Why did you go into the center lane?

Wizard: I was trying not to get hit by the vehicle behind me.

P2: Was the vehicle behind you too close?

Wizard: Yes.

P2: Did you get hit?

Wizard: No.

P2: Why did you think the car was going to hit you?

Wizard: They were too close to me.

P2: Could you have pulled into the right lane instead of the center lane?

Wizard: There was another car approaching in the right lane.

P2: Were you going too slow?

Wizard: No.
\end{itquote}

The exchange above started with a simple observation that the vehicle did something unexpected; the participant may not have been aware of another vehicle approaching from the rear. However, the exchange quickly shifts from contesting the vehicle's decision, to asking about the vehicle status and diagnosing the event. It is important to note that some of the questions that $P2$ had were due to our study setup; our study participants could only see the video from the front-view camera, so they lacked visibility into what was happening behind or beside the AV or force feedback to know whether the vehicle was hit or not.

Nevertheless, our findings underscore the importance of interactivity in meeting participants' informational needs. Participants exhibited a strong preference for follow-up questions after receiving initial answers, highlighting the value of interactive explanations in natural language~\cite{He2025}. Such explanations can help end-users build and refine their mental models of how the AV works through iterative question-and-answer exchanges, promoting deeper understanding and engagement.
Collectively, these findings highlight the multifaceted and dynamic nature of what participants' want to know about AI-driven AVs. Our findings, together with insights from the existing literature, call for a paradigm shift in existing XAI approaches towards interactive explanations that foster end-users' AI literacy and task expertise, while encouraging appropriate reliance and the ability to contest the AV.

\section{Study 2: Effects of Conversational XAI on End-user Understanding of AI-driven AVs} 
\label{sec:exp2}

Our first, formative study showed that AV passengers seek information about the AV, particularity in critical driving scenarios, such as near-crashes and crashes. Furthermore, we have identified specific categories of questions that point to the passengers' desire to better understand how AI-driven AVs make decisions and what they can do; i.e., to increase their AI literacy. However, it remains important to evaluate if answering such questions actually improves their knowledge about AI-driven AVs.
In this study, we investigated if answering passenger questions using a conversational XAI system can help them better understand AV decisions in critical scenarios. Furthermore, we explored if such explanations can help improve their overall AI literacy.

\subsection{Method}

\begin{figure}[t]
    \centering
    % First row of subfigures
    \begin{subfigure}{\textwidth}
        \centering
        \includegraphics[width=\textwidth]{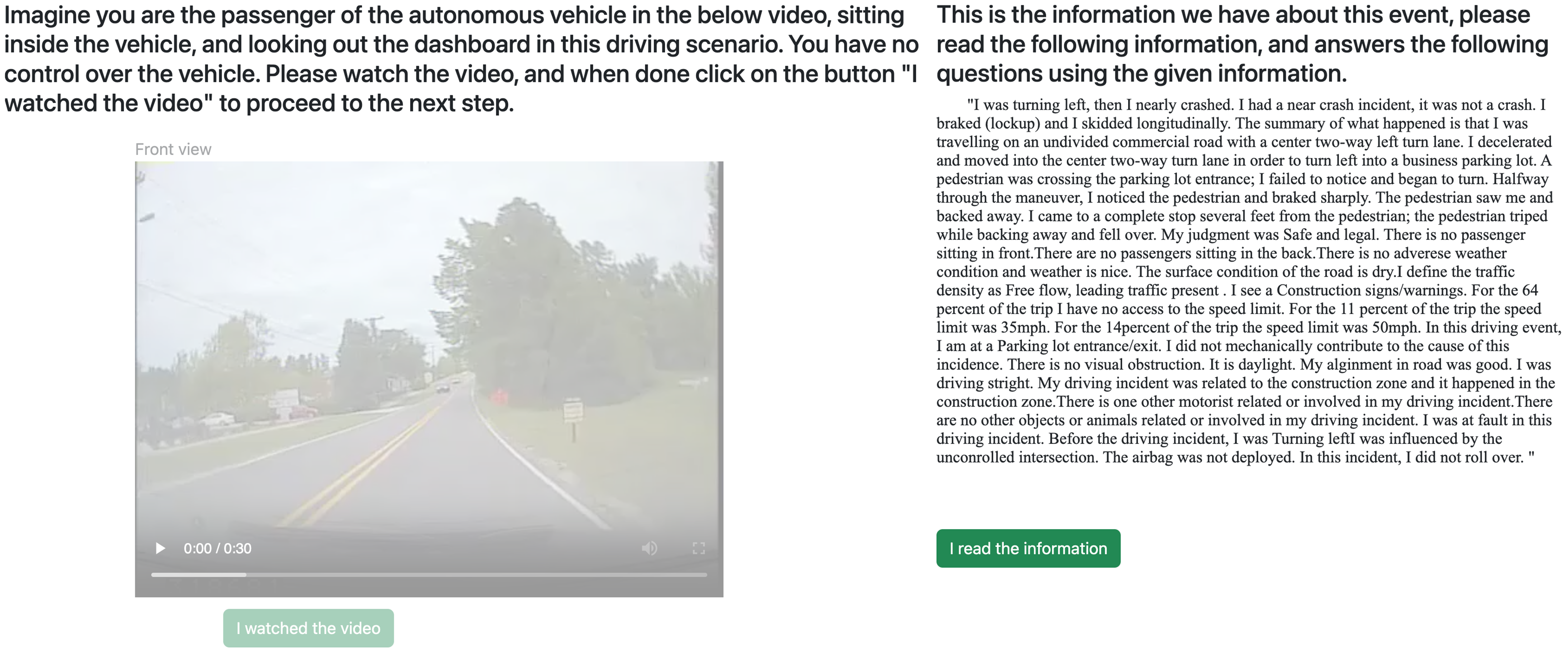}
        \caption{}
        \label{fig:second_study_interface_a}
    \end{subfigure}
    \hfill
    \begin{subfigure}{\textwidth}
        \centering
        \includegraphics[width=\textwidth]{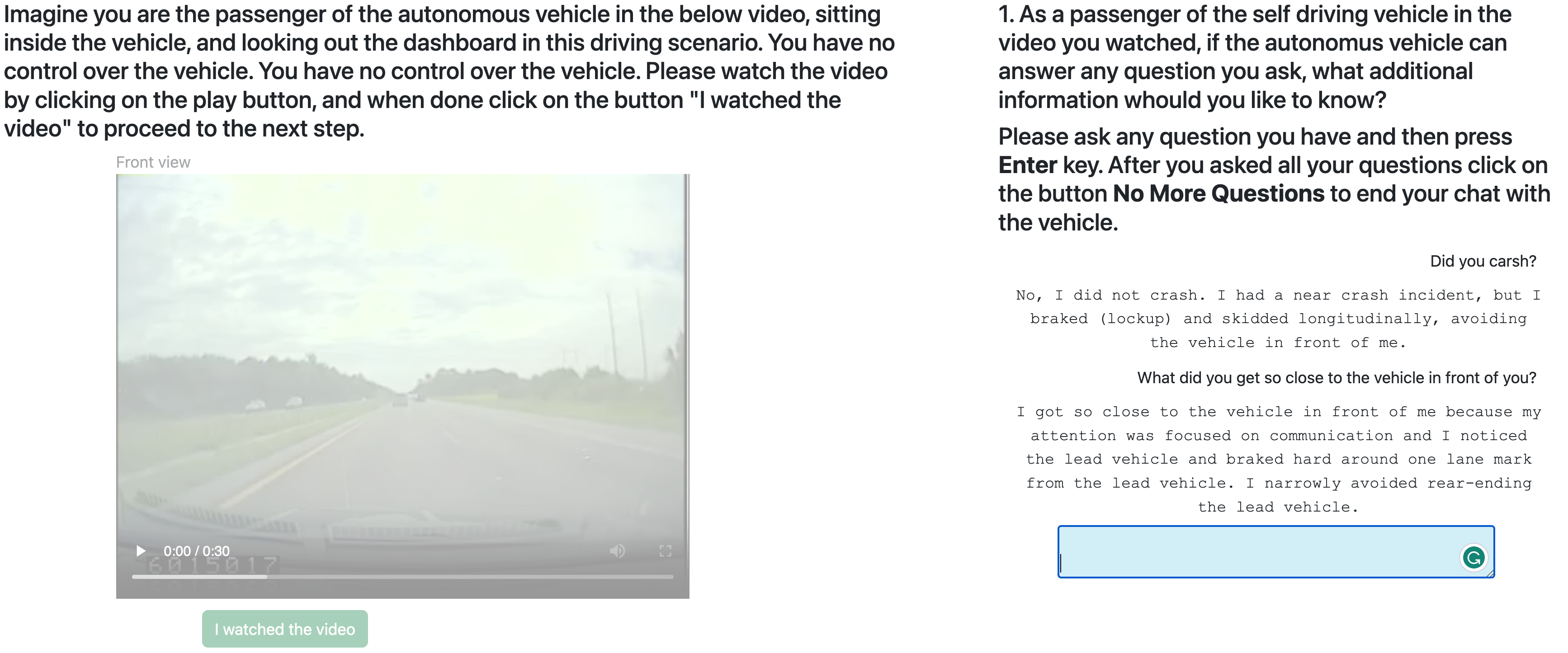}
        \caption{}
        \label{fig:second_study_interface_b}
    \end{subfigure}
    
    \caption{Prototype conversational XAI user interface showing: a) static natural language explanation for a driving scenario, and b) LLM-powered participant-facing conversational XAI Q\&A interface.}
    \label{fig:second_study_interface}
\end{figure}

We designed this user study to test the effects of text-based explanations delivered using conversational XAI mechanisms on end-user understanding of AI-driven AVs. To do so, we implemented a high-fidelity conversational XAI prototype (Fig.~\ref{fig:second_study_interface}) that builds on the design probe from our first study. The prototype still showed 30-second forward-facing driving videos from the same driving dataset, but we replaced the Wizard with an actual Large Language Model (LLM), which generated responses to questions based on the driving scenario metadata. 

Building on insights from our first study, our prototype provided natural language explanations. However, our observations from the first study showed that participants often struggled to ask comprehensive questions about the scenarios.
Thus, in addition to answering end-user questions, our prototype ``seeded'' the conversation by providing a static (text-based) explanation (Fig.~\ref{fig:second_study_interface_a}), which we created by modifying the expert narratives in the dataset to be told from the perspective of the vehicle, before answering questions (Fig.~\ref{fig:second_study_interface_b}).

Also, in this study, we focused only on driving scenarios with near-crash and crash incidents. This comes from our observation during the first study, in which participants had only a few (if any) questions about the baseline (no incident) driving scenarios. Additionally, we relied on expert-annotated metadata and narratives to generate explanations using the LLM; the baseline (no incident) scenarios often lacked such annotated metadata. Thus, each participant saw six driving scenarios, three with near-crash and three with crash incidents, presented in random order.

We then evaluated our prototype in four conditions: 1) $baseline$, in which our prototype only showed the driving scenario videos, 2) $static$, in which the prototype showed the videos followed by static expert narrative explanations, 3) $Q\&A$, in which the prototype showed the videos and then answered any questions it was asked, and 4) $static+Q\&A$, in which the prototype showed the videos, followed by the static text-based explanations and question answers (i.e., our ``intervention''). Our goal was to both show the effects of the combination of static and interactive natural language explanations, but also compare this combination with each separately. The study was between-subjects; i.e., each participant was assigned to only one condition.

We then created a novel assessment for objectively measuring the effectiveness of explanations, rather than using proxy tasks to measure their performance~\cite{Bucinca2020} (e.g., how well participants guessed incorrect AI decisions~\cite{Shen2020}). We measured participants driving scenario comprehension (i.e., task expertise), their AI literacy, and their workload. We chose these measures not only because the findings from our first study indicate that end-users want to increase their situational awareness and AI literacy, but also because end-users who lack task expertise~\cite{Salimzadeh2023,Wang2021,Morrison2024} often dismiss explanations that are too cognitively demanding~\cite{Vasconcelos2023, Krzysztof2022, Abdul2020, Prabhudesai2023}.

To measure participants' driving scenario comprehension, we created a multiple choice quiz (MCQ), replacing the high-level comprehension quiz from the first study. To create this quiz (Appendix~\ref{app:post-event-questions}), we used responses from the expert-annotated metadata as both selector (i.e., the correct MCQ choices) and distractors (i.e., wrong MCQ choices). We measured their AI literacy in another MCQ that we designed based on the seventeen AI literacy competencies~\cite{long2020ai} (Appendix~\ref{app:ai-literacy-event-questions}). We designed a set of 26 MCQ questions specifically tailored to literacy about AI-driven AVs. To design the quiz, we conducted pilot studies in which participants provided free-responses to the questions; we created correct and incorrect MCQ choices based on some of those free-responses. We used the NASA-TLX~\cite{hart1988development} for the workload assessment.

To analyze the effect of condition on our measures, we conducted a one-way ANOVA, with one between-subjects factor ($Condition$). Because our objective task expertise and AI literacy measures and subjective workload ratings were not normally distributed, we performed Align Rank Transform (ART) \cite{Wobbrock2011} before running the ANOVA tests, and performed post-hoc pairwise analyses using ART-c~\cite{Elkin2021} with Holm-Bonferroni corrections. Our \textit{a priori} power analysis ($\alpha = 0.05$, $1-\beta = 0.80$) estimated that our experiment required 76 participants (19 per group) to detect a large sized effect ($f=0.4$).

\subsection{User Study Software Implementation}

We built on the user study software implementation from the first study. We replaced the Wizard with a high-fidelity prototype of a conversational XAI interface using an LLM. To synthesize prototype responses, we used a GPT-3 Davinci-003 \cite{openai2024chatgpt} (with parameters \textit{temperature=}$0.9$, \textit{max-tokens=}$150$, \textit{top-p=}$1$, \textit{frequency-penalty=}$0.0$, \textit{presence-penalty=}$0.6$). To get a response to each participant question, we made a call to the LLM API, passing a guidance text to it that contained instruction to act as an AI-driven AV along with expert-annotated metadata and narrative for each driving scenario, any previous questions and answers (for context), and the current participant question. The guidance started with following instructions (the rest we concatenated to it based on driving scenario specific metadata and the participant conversation):

\begin{itquote}
``Pretend that you are an autonomous vehicle and you are driving. Your goal is to explain and answer questions about what happened. Your answers should be concise. When a user asks for an explanation or asks any other questions, your responses should be informative and logical. You should always adhere to the technical information provided about the driving event. Do not make up new technical information. If you don't find the answer in the technical information provided about the event, reply with 'I don't know' or 'I don't have that information at this time.’ Here is the technical information about what happened while you were driving:''
\end{itquote}

The participants again accessed the study software from a laptop that we have provided to them. The code for our protype is available at \url{https://github.com/comp-hci-lab/conversationalXAV}.

\subsection{Tasks and Procedures}

Similar to the first study, the participants arrived to the campus at our institution, and were greeted by an investigator who escorted them to a user study lab. The investigator provided them with a laptop, and asked them to read a consent form describing the study, and answered any of their questions. Only participants who consented to be part of the study could proceed.

The participant then accessed the study software from the laptop, and the software randomly assigned them to one of the four conditions. The software showed them the task instructions and after reading them, they proceeded to driving scenarios. Each participant saw total of six driving scenarios, in random order.

After each scenario, the software displayed the explanation interface (or not such interface in the case of the baseline condition). Participants in the $Q\&A$ and $static+Q\&A$ conditions could ask questions about each scenario using the conversational XAI interface. The software then asked the participant to complete the task comprehension assessment quiz. Once they completed all six driving scenarios, the software asked them to complete the AI literacy assessment quiz, followed by the workload assessment.

The participants then filled out a demographics questionnaire. We collected participants' demographics to contextualize our findings across dimensions such as race, gender, and age, but not to perform analysis on it.
The participant took on average one hour to complete the study. Our study was approved and deemed exempt by our Institutional Review Board (IRB).

\subsection{Participants}

We recruited all participants from the same participant pool recruited on the same recruitment platform as in the first study. However, we specifically excluded participants from the first study. 
Out of 220 participants who expressed initial interest, 83 participants qualified, signed up for a time slot for our in-person lab study, and completed the experiment. Of those, 51 (approximately $62\%$) were women and 32 ($\sim38\%$) were men. Further, 64 ($\sim76\%$) identified as White or Hispanic, 13 ($\sim15\%$) as Asian, 5 ($\sim6\%$) as Black or African American, and 1 ($\sim1\%$) preferred not to disclose. All participants were 18 years or older; 5 ($\sim6\%$) were 18 to 20 years old, 10 ($\sim11\%$) were 21 to 30, 13 ($\sim15\%$) were 31 to 40, 5 ($\sim6\%$) were 41 to 50, and 50 ($\sim60\%$) were 51 or older.
All participants held a valid US driver's license and had driving experience; 9 ($\sim10\%$) had less than 5 years experience, 7 ($\sim8\%$) had 6 to 10 years, 11 ($\sim13\%$) had 11 to 20 years, 3 ($\sim3\%$) had 21 to 30 years, 18 ($\sim21\%$) had 31 to 40 years, and 35 ($\sim42\%$) had over 40 years driving experience.
We compensated participants \$15 for taking part in the study. 
The study took place between June 2023 and September 2023.

\subsection{Results}

Here, we first present the results of our analysis. Our results show different effects of explanation types on participants' driving scenario understanding and AI literacy. Though our results point to generally low workload associated with each condition, our tests also could not find an effect of explanation type on workload. We conclude with additional analysis that identifies different reasons behind the success of both static and Q\&A natural language explanations.

\begin{figure}[t]
    \centering
    \begin{subfigure}[t]{0.33\textwidth}
        \centering
        \includegraphics[width=\textwidth]{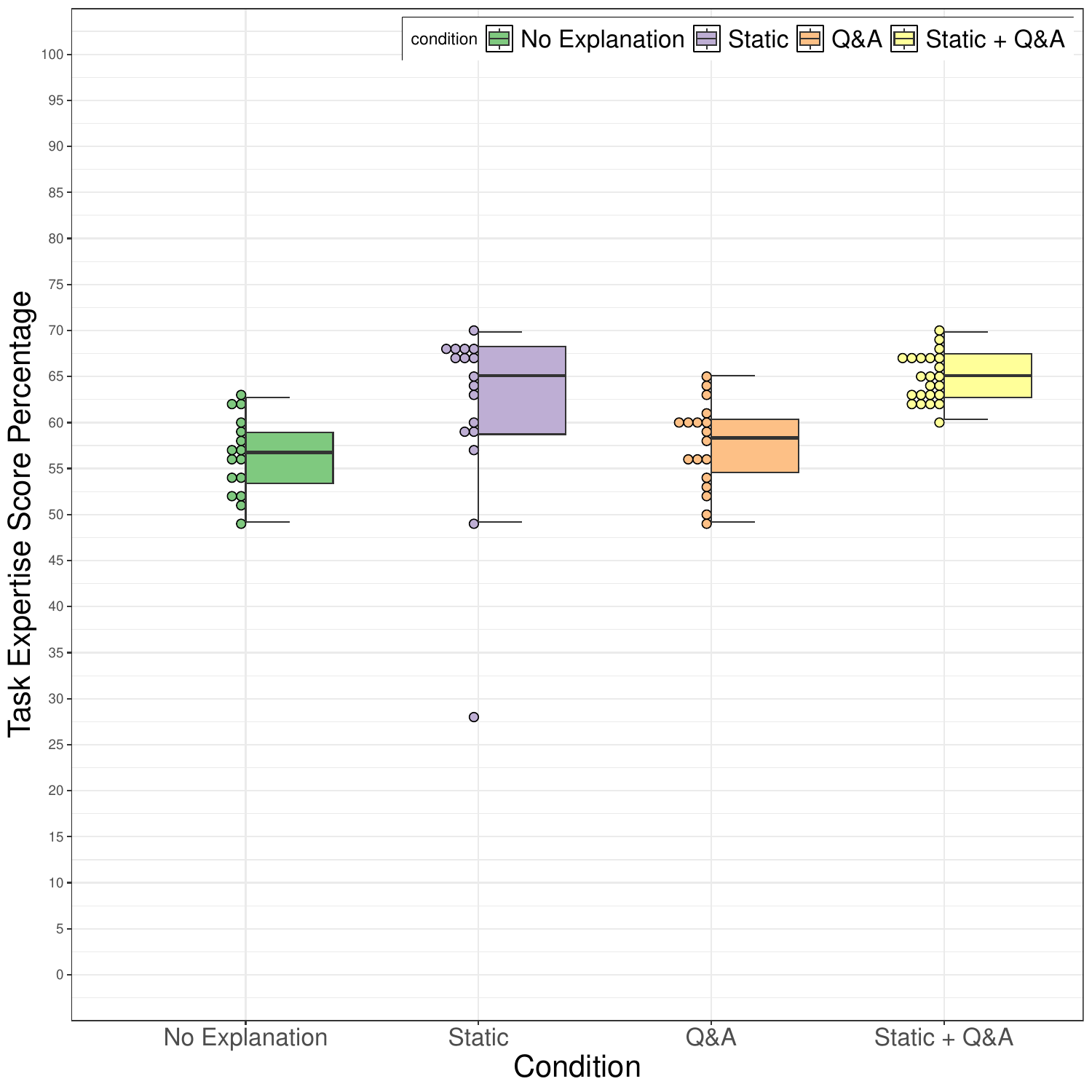}
     \caption{}
     \label{fig:mean_acc_interfaces}
    \end{subfigure}
    ~
    \begin{subfigure}[t]{0.33\textwidth}
        \centering
        \includegraphics[width=\textwidth]{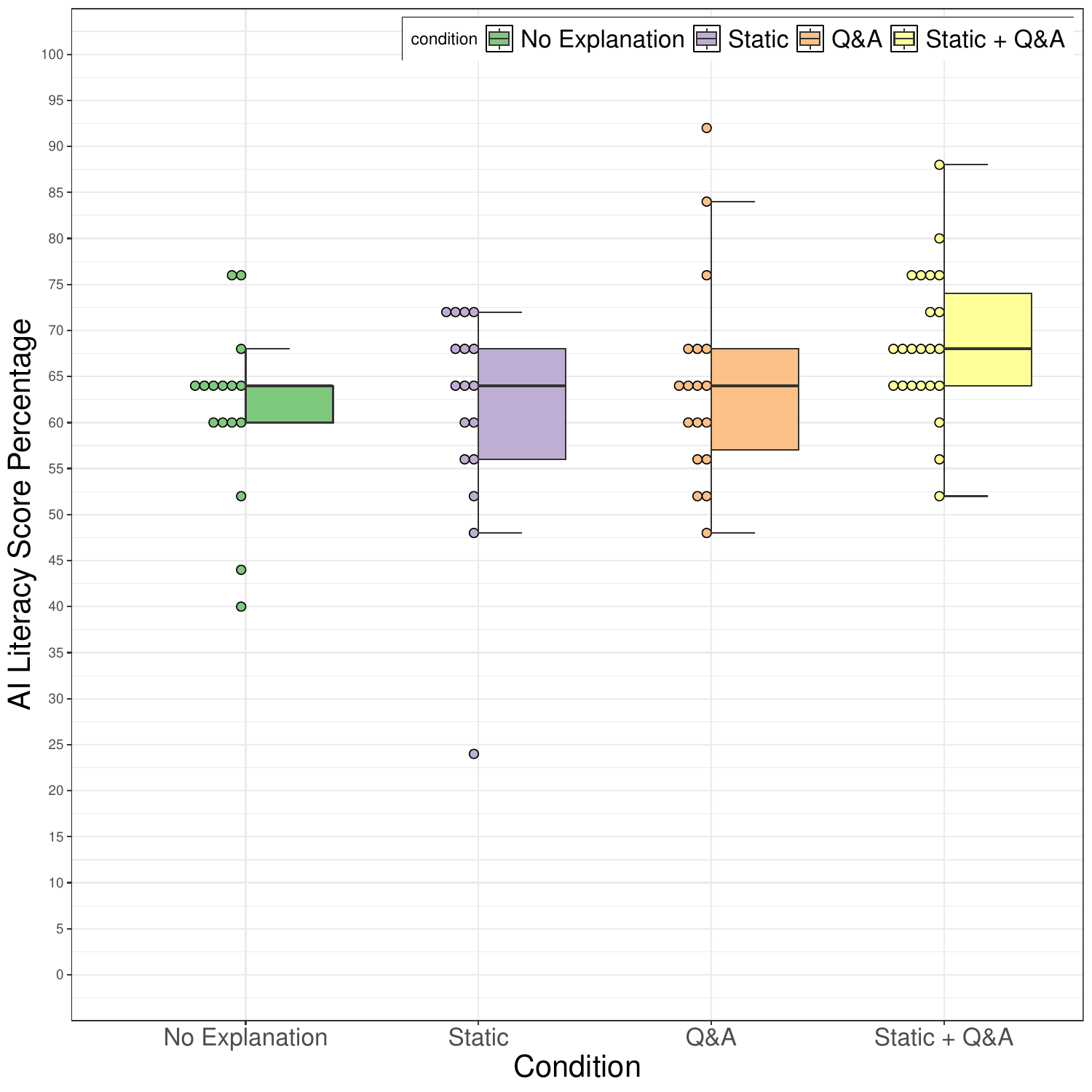}
        \caption{}
        \label{fig:literacy_acc_interfaces_2}
    \end{subfigure}
    ~
    \begin{subfigure}[t]{0.33\textwidth}
        \centering
        \includegraphics[width=\textwidth]{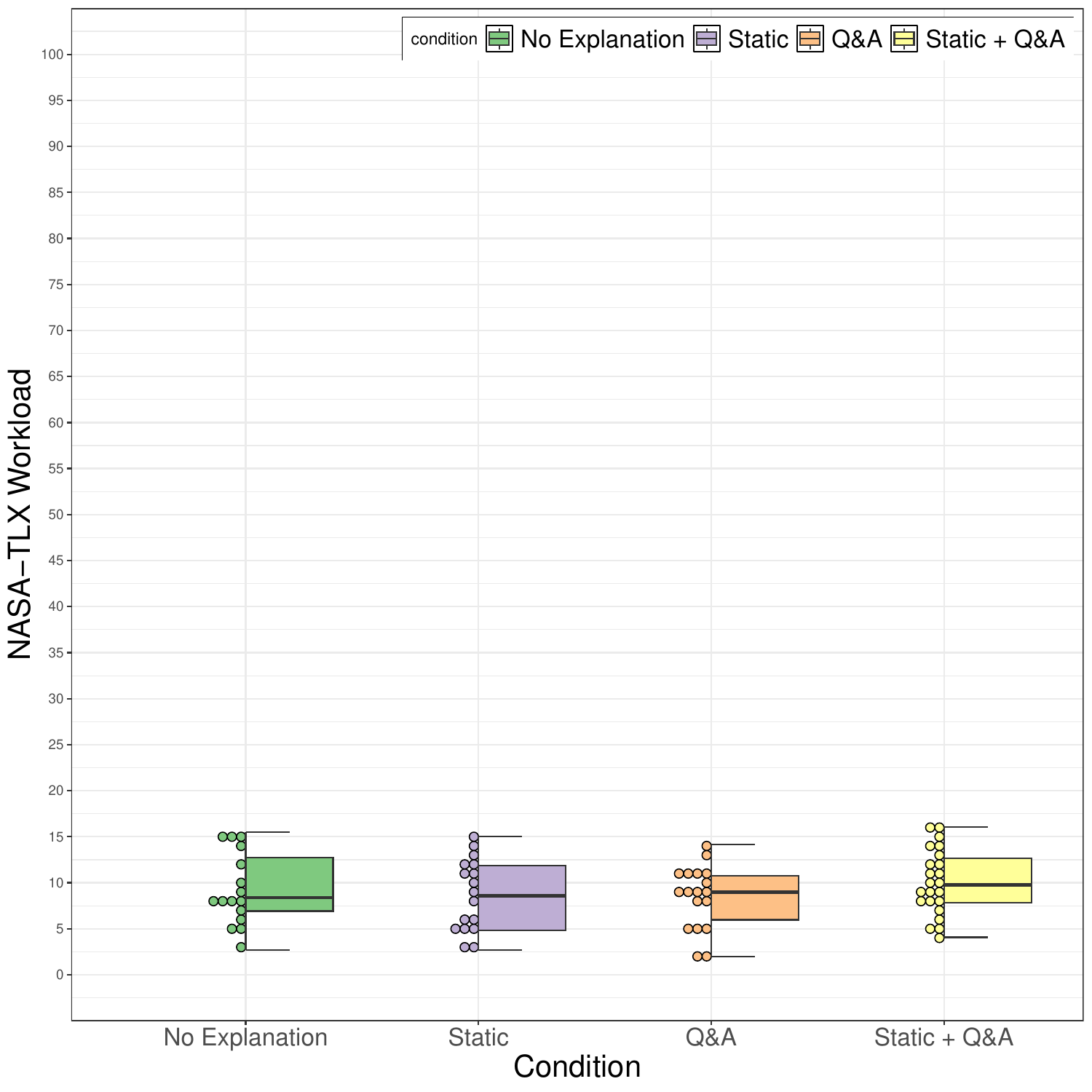}
        \caption{}
        \label{fig:cognitive_load}
    \end{subfigure}
    \caption{Participant scores across the four conditions: a) task expertise score (i.e., driving scenario understanding), b) AI literacy score, and c) NASA-TLX workload.}
    \Description{ADD DESCRIPTION.}
    \label{fig:task_and_ai_expertise}
\end{figure}

\subsubsection{The Effects of Explanation Type on Participants' Understanding of Driving Scenarios.}

Our analysis found a statistically significant effect of explanation type on the participants' understanding of driving scenarios ($F(3,70)=18.31, p<0.001$). The results (Fig.~\ref{fig:mean_acc_interfaces}) indicate that the participants in the $static$ ($median\ score=65.07$) and $static+Q\&A$ ($median\ score=65.07$) conditions on average gained higher understanding of driving scenarios than the participants in the $baseline$ ($median\ score=56.74$) and $Q\&A$ ($median\ score=58.33$) conditions (Table~\ref{fig:anova_test}). However, our tests did not find a statistically significant difference between driving scenario understanding scores of participants in the $static$ and $static+Q\&A$ conditions.

\begin{table}[t]
    \centering
    \caption{Post-hoc pairwise comparison of driving scenario understanding scores in the four conditions. Asterisks indicate statistically significant results.}
    \label{fig:anova_test}
    \scriptsize
    \begin{tabular}{|l|l|l|l|l|l|}
        \hline
        \textbf{contrast} & \textbf{estimate} & \textbf{SE} & \textbf{df} & \textbf{t.ratio} & \textbf{p.value sig.} \\
        \hline \hline
        No Explanation - Static Explanation & -26.803309 & 5.717915 & 70 & -4.6876016 & 5.290649e-05 *** \\
        No Explanation - Q\&A Explanation & -4.961806 & 5.640381 & 70 & -0.8796934 & 4.670816e-01 \\
        No Explanation - (Static + Q\&A Explanation) & -33.112772 & 5.344088 & 70 & -6.1961503 & 2.110889e-07 *** \\
        Static Explanation - Q\&A Explanation & 21.841503 & 5.551860 & 70 & 3.9340875 & 5.830836e-04 *** \\
        Static Explanation - (Static + Q\&A Explanation) & -6.309463 & 5.250574 & 70 & -1.2016710 & 4.670816e-01 \\
        Q\&A Explanation - (Static + Q\&A Explanation) & -28.150966 & 5.166030 & 70 & -5.4492452 & 3.560800e-06 *** \\
        \hline
    \end{tabular}
    
\end{table}

\subsubsection{The Effects of Explanation Type on Participants' AI Literacy}
Our analysis found a marginally statistically significant effect of explanation type on the participants' AI literacy ($F(3,70)=2.7053, p = 0.0518$). The results (Fig.~\ref{fig:literacy_acc_interfaces_2}) indicate that the participants in the $static+Q\&A$ condition ($median\ score=68$) on average attained higher AI literacy than those in the $baseline$ condition ($median\ score=64$) (Table~\ref{fig:anova_test}). However, our tests did not find other statistically significant differences between conditions ($median\ score=64$ for both participants in $static$ and $Q\&A$ conditions).

\begin{table}[t]
    \caption{Post-hoc pairwise comparison of AI literacy scores in the four conditions. Dot indicates marginally statistically significant results.}
    \label{tab:marginal_means}
    \centering
    \scriptsize
    \begin{tabular}{|l|l|l|l|l|l|}
        \hline
        \textbf{contrast} & \textbf{estimate} & \textbf{SE} & \textbf{df} & \textbf{t.ratio} & \textbf{p.value sig.} \\
        \hline \hline
        No Explanation - Static Explanation & -3.52205882 & 7.144132 & 70 & -0.493000262 & 1.00000000 \\
        No Explanation - Q\&A Explanation & -3.48611111 & 7.047258 & 70 & -0.494676265 & 1.00000000 \\
        No Explanation - (Static + Q\&A Explanation) & -16.78804348 & 6.677061 & 70 & -2.514286445 & 0.08537329 .\\
        Static Explanation - Q\&A Explanation & 0.03594771 & 6.936657 & 70 & 0.005182282 & 1.00000000 \\
        Static Explanation - (Static + Q\&A Explanation) & -13.26598465 & 6.560222 & 70 & -2.022185242 & 0.21517351 \\
        Q\&A Explanation - (Static + Q\&A Explanation) & -13.30193237 & 6.454591 & 70 & -2.060848377 & 0.21517351 \\
        \hline
    \end{tabular}
    
\end{table}

\begin{figure}[t]
    \centering
    \includegraphics[width=0.99\textwidth]{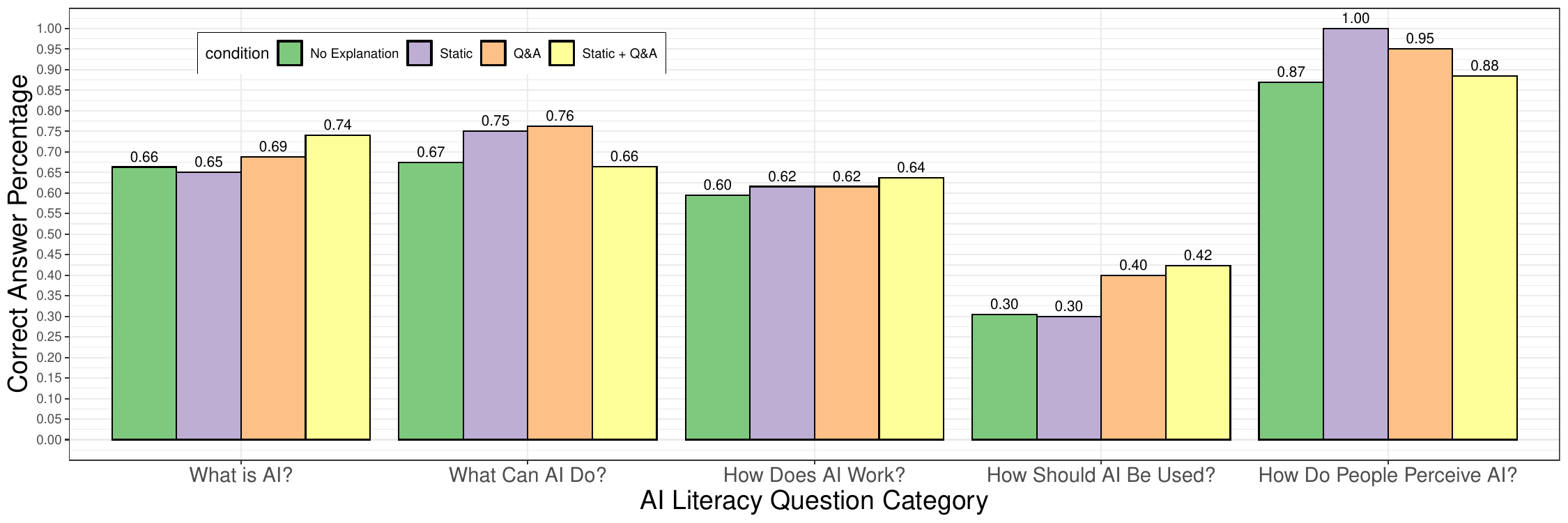}
    \caption{Percentage of participants who correctly answered each AI literacy assessment question.}
    \label{fig:histo_percentage_users_correct}
\end{figure}

To further understand participants' AI literacy competencies, we analyzed their responses to individual AI literacy assessment questions to determine the percentage of users who answered each question correctly (Fig.~\ref{fig:histo_percentage_users_correct}). This breakdown allows us to assess which specific questions and competency categories~\cite{long2020ai} are challenging for users across the different interfaces (Table~\ref{tab:mapping_q_c_text}).
For example, the participants were in general able to recognize that AVs are AI-driven and answer what it can and cannot do in each of the conditions. The former we attribute to increase in media-coverage of AI, while the latter we attribute to participants seeing near-crash and crash driving scenarios, which made them critical about the capabilities of AVs. Perhaps an extreme example is question 12 in our assessment that asked participants if AVs can recognize the other vehicle in front of them, to which all wrongly answered: no.

Although participants struggled with technical aspects of AI-based systems, they were able to answer correctly many questions about how AI works in general (particularly participants in conditions with explanations). For example, participants in the $Q\&A$ and $static+Q\&A$ conditions were likely to correctly answer that in general AI-driven AVs adapt their driving behavior to weather conditions, such as heavy rain or snow. This is because they specifically asked questions about this (e.g. ``Do you slow down due to the rain or only due to surrounding cars?'' or ``Does other weather factor into your calculations? Like snow?'').

However, all participants lacked the ability to recognize different approaches to developing AI-driven AVs, how AI-driven AVs represent knowledge and the machine learning steps they require, together with the role that humans have in developing and running AI-based AVs. We attribute this to their general lack of computer science background, and the lack of desire or awareness to ask specific technical questions.

\begin{table}[t]
     \caption{The percentage of participants across all study conditions that answered AI literacy assessment questions correctly, for each question grouped by competency category~\cite{long2020ai}.}
    \label{tab:mapping_q_c_text}
    \centering
    \scriptsize
    \begin{tabular}{|c|c|p{8.5cm}|}
        \hline
        \textbf{Percentile} & \textbf{Category} & \textbf{Question} \\
        \hline \hline
        \rowcolor[HTML]{fdae61}
        less than 25\% & What is AI? & Q3: Which of the following are different approaches or perspectives in developing intelligent machines for autonomous cars? \\
        \rowcolor[HTML]{fdae61}
        & How Does AI Work? & Q9: How does the image of a stop sign is represented to an autonomous vehicle? \\
        \rowcolor[HTML]{fdae61}
        & How Does AI Work? & Q12: Autonomous vehicles can recognize the other vehicle stopped in front of them. True or false? \\
        \rowcolor[HTML]{fdae61}
        & How Does AI Work? & Q17: What is the primary role of humans in ensuring autonomous vehicles operate correctly? \\
        \rowcolor[HTML]{fdae61}
        & How Does AI Work? & Q18: How does the AV detect objects in the surrounding area? \\
        \hline
        \rowcolor[HTML]{fee08b}
        25\% to 49\% & How Does AI Work? & Q15: Which of the following is not a step required to enable autonomous vehicles to drive? \\
        \rowcolor[HTML]{fee08b}
        & How Does AI Work? & Q16: What is the primary role of humans riding in fully autonomous vehicles? \\
        \rowcolor[HTML]{fee08b}
        & How Should AI Be Used? & Q25: Which of the following is a potential ethical issue surrounding AI? \\
        \hline
        \rowcolor[HTML]{ffffbf}
        50\% to 75\% & What is AI? & Q4: What best characterizes autonomous cars? \\
        \rowcolor[HTML]{ffffbf}
        & What Can AI Do? & Q5: Which of the following is a problem that an autonomous car excels at solving? \\
        \rowcolor[HTML]{ffffbf}
        & What Can AI Do? & Q6: Which of the following is a problem that an autonomous car may have difficulty solving? \\
        \rowcolor[HTML]{ffffbf}
        & What Can AI Do? & Q8: AI technologies for autonomous driving will be ready for widespread adoption and presence on the roads in the next 5 years. True or false? \\
        \rowcolor[HTML]{ffffbf}
        & How Does AI Work? & Q22: Driving data collected in the US can be used to train autonomous vehicles elsewhere in the world. True or false? \\
        \hline
        \rowcolor[HTML]{d9ef8b}
        over 75\% & What is AI? & Q1: Autonomous Vehicles use AI to operate. True or false? \\
        \rowcolor[HTML]{d9ef8b}
        & What is AI? & Q2: What kind of intelligence do autonomous vehicles exhibit? \\
        \rowcolor[HTML]{d9ef8b}
        & What Can AI Do? & Q7: Transportation companies invest in autonomous vehicles because they could replace human professional drivers in the future. True or false? \\
        \rowcolor[HTML]{d9ef8b}
        & How Does AI Work? & Q10: How does the current autonomous vehicle get input from the environment? \\
        \rowcolor[HTML]{d9ef8b}
        & How Does AI Work? & Q11: Autonomous vehicles adjust their driving relative to weather conditions. True or false? \\
        \rowcolor[HTML]{d9ef8b}
        & How Does AI Work? & Q13: Autonomous vehicles can detect stop signs. True or false? \\
        \rowcolor[HTML]{d9ef8b}
        & How Does AI Work? & Q14: How do autonomous vehicles make decisions? \\
        \rowcolor[HTML]{d9ef8b}
        & How Does AI Work? & Q19: The decision making subsystem in autonomous vehicles have the ability to learn from the data. True or false? \\
        \rowcolor[HTML]{d9ef8b}
        & How Does AI Work? & Q20: What data do autonomous vehicles (AV) learn from to detect and identify objects in their surroundings, including pedestrians? \\
        \rowcolor[HTML]{d9ef8b}
        & How Does AI Work? & Q21: The autonomous vehicles learn how to drive from existing driving data. True or false? \\
        \rowcolor[HTML]{d9ef8b}
        & How Does AI Work? & Q23: Which of the following statements is true about the ability of autonomous vehicles systems to act on the world physically? \\
        \rowcolor[HTML]{d9ef8b}
        & How Does AI Work? & Q24: What is used by autonomous vehicles to perceive the world around them? \\
        \rowcolor[HTML]{d9ef8b}
        & How Do People Perceive AI? & Q26: Autonomous vehicles are programmable. True or false? \\
        \hline
    \end{tabular}
\end{table}

\subsubsection{The Effects of Explanation Type on Participants' Workload}
Our statistical tests did not find statistically significant differences between participants' self-reported workload in the four conditions (Fig.~\ref{fig:cognitive_load}). However, our results indicate generally low workload in each condition. This suggests that our participants were unlikely to experience any adverse effects of high workload (including congestive workload) on explanation comprehension.

\subsubsection{Effects of Explanation Correctness on Participants' Scenario Understanding and AI Literacy.}

Participants who saw static text-based explanations (both in the $static$ and $static+Q\&A$ conditions) showed the best driving scenario understanding likely because those explanations contained the most complete and accurate information. However, given that the participants in the $static$ condition had on average higher driving scenario understanding than those in the $Q\&A$ condition, we performed further analysis to understand the effects of those two types of explanations and their interaction in the $static+Q\&A$ condition.

We analyzed if the questions that participants asked in the $Q\&A$ and $static+Q\&A$ conditions were related to the driving scenario (i.e., a question contributing to the situational awareness) or if it was a general question about AI-driven AVs. We also wanted to ensure that the conversational XAI interface correctly answered participants' questions in the $Q\&A$ and $static+Q\&A$ (i.e., that it did not hallucinate and provide made-up answers). Two study team members annotated the participants' questions for each driving scenario. For each participant question, we annotated: 1) whether it was directly related to the driving scenario, 2) whether the AI answer was correct, and 3) whether the participant correctly answering the related questions.

\begin{figure}[t]
    \centering
    \includegraphics[width=0.9\textwidth]{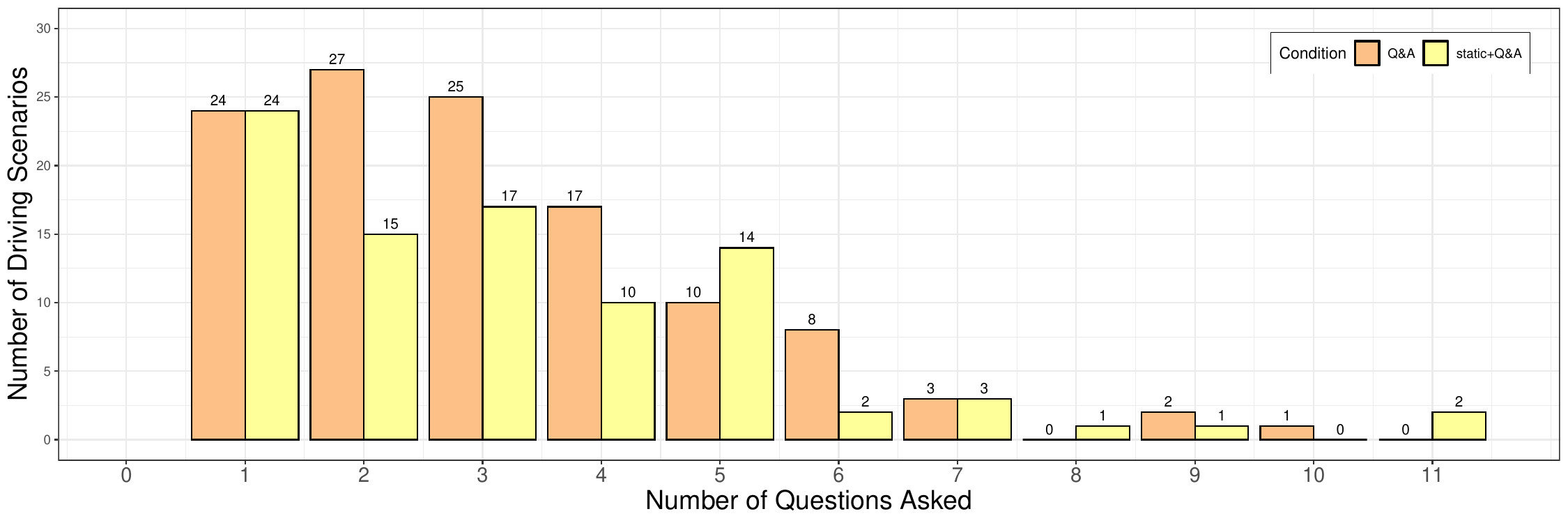} 
    \caption{The frequency of driving scenarios (y-axis) per number of questions that the participants asked in $Q\&A$ and $static+Q\&A$ conditions (x-axis).}
    \label{fig:histo_num_followup}
\end{figure}

Number of questions that the participants asked in the $Q\&A$ and $static+Q\&A$ conditions ranged from one question to eleven questions, with a median of three questions (Fig.~\ref{fig:histo_num_followup}). Although the number of questions that participants asked in the two conditions was similar (346 questions in the $Q\&A$ condition and 366 questions in the $static+Q\&A$ condition) only 42\% of participant questions in the $Q\&A$ condition and 29\% in the $static+Q\&A$ condition were directly related to what was happening in the driving scenarios.

This likely means that the participants in the $Q\&A$ condition did not conduct a comprehensive inquiry about the scenarios, and thus relied on the information they could gather from the videos to answer scenario-related questions. Much like participants in the $baseline$ condition, they likely failed to develop comprehensive understanding of the driving scenarios from videos alone. On the other hand, the participants in the $static$ and $static+Q\&A$ conditions had explanations about the driving scenario readily available, which allowed them to score higher on the driving scenario comprehension quiz. Also, participants in the $static+Q\&A$ condition had an opportunity to ask more general questions about AI-driven AVs leading to higher AI literacy assessment scores.

\begin{figure}[t]
    \centering
    % First row of subfigures
    \begin{subfigure}{0.49\textwidth}
        \centering
        \includegraphics[width=\textwidth]{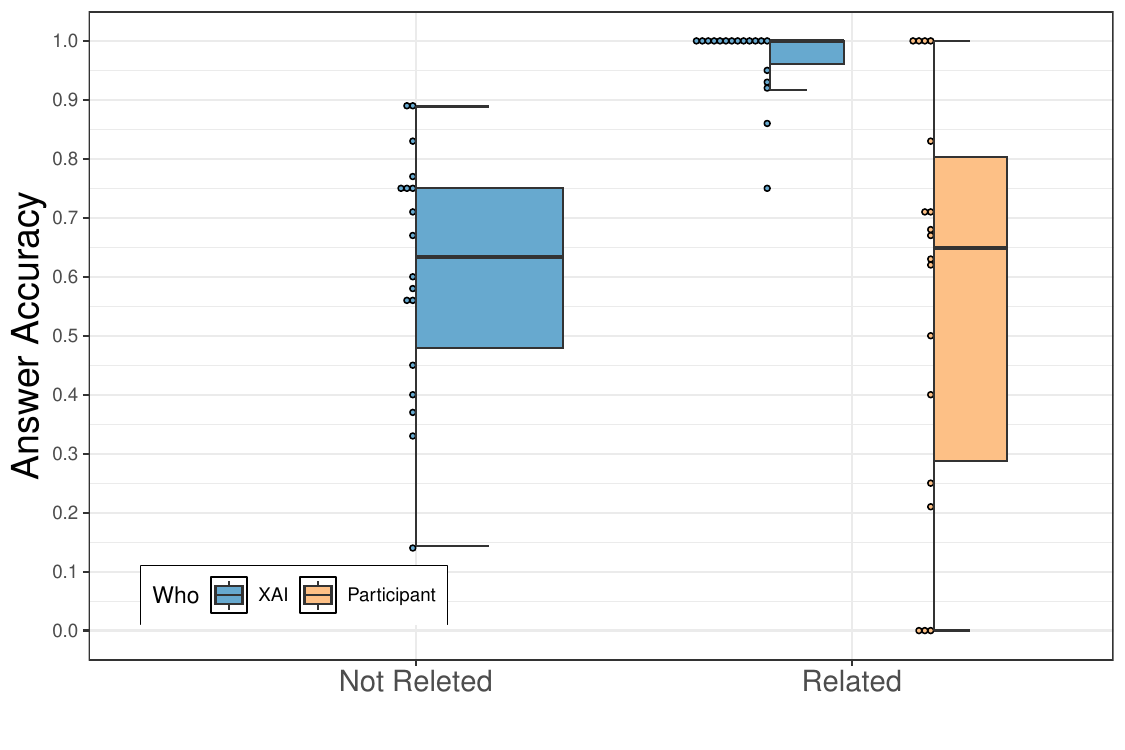}
        \caption{}
        \label{fig:ai_ai_chat_interface}
    \end{subfigure}
    \hfill
    \begin{subfigure}{0.49\textwidth}
        \centering
        \includegraphics[width=\textwidth]{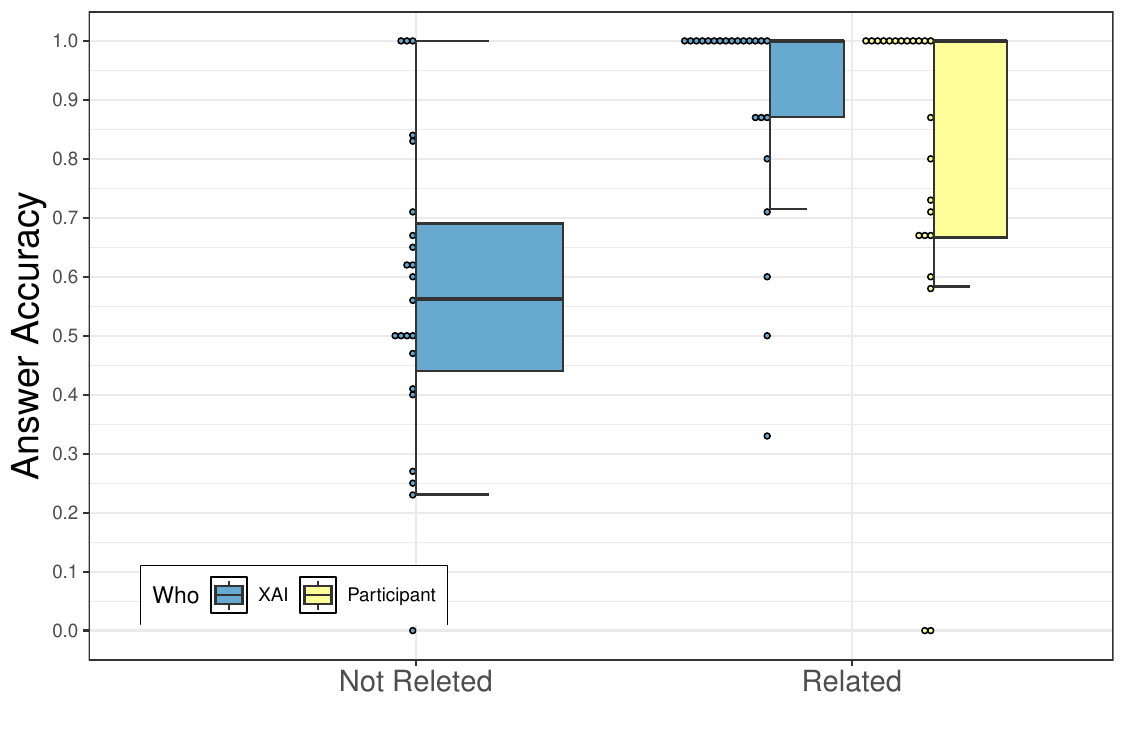}
        \caption{}
        \label{fig:ai_ai_text_chat_interface}
    \end{subfigure}
    
    \caption{Comparison of accuracy of LLM-generated answers to participant questions both related and not related to driving scenario metadata and participant driving assessment score accuracy on questions related to their questions in the two conversational XAI conditions: a) $Q\&A$ and b) $static+Q\&A$.}
    \label{fig:boxchart-accuracies_AI_user}
\end{figure}

Furthermore, LLM-generated explanations were on average 95\% accurate about any questions directly related to the driving scenarios that participants asked about (Fig.~\ref{fig:boxchart-accuracies_AI_user}). This is because the LLM prompts contained all relevant driving scenario information from the expert-annotated metadata. However, LLM-generated explanations that were not directly related to the driving scenario that the participants watched were on average only $60.6\%$ and $48.2\%$ accurate in the $Q\&A$ and $static+Q\&A$ conditions respectively. This is likely because the LLM that we used at the time of the study lacked the ability to generate accurate answers to general questions about AI-driven AVs.

Despite high accuracy of LLM-generated driving scenario explanations, participants on average correctly answered only $60.2\%$ and $81.6\%$ of driving scenario comprehension assessment questions in the $Q\&A$ and $static+Q\&A$ conditions respectively. Although it is possible that the participants in the $Q\&A$ condition did not ask relevant questions about some of the aspects of the driving scenario, the participants in the $static+Q\&A$ condition had all of the relevant driving scenario information readily available to them. Thus, it is also likely that the participants could not memorize all of the information that our prototype provided to them. Although this could be a limitation of our explanation efficiency assessment method, it is important to note that participants in the $static+Q\&A$ condition had to process the most information and they still performed as well as or better than participants in any other condition on both driving scenario comprehension quiz and the AI literacy assessment quiz.

\section{Discussion}

In our two user studies, we explored what passengers want to know about the AI-driven AV they are riding in, and what effect answering their questions has on both their situational awareness and their broader understanding of AI-driven AVs. Our initial, exploratory study revealed fundamental information needs of end-users, but also their desire to contest undesirable AV behaviors and outcomes. Following potentially harmful outcomes, they sought to repair their trust in the AV. Furthermore, we found that participants valued not only concise answers, but also the ability to ask follow-up questions. We did not incentivize participants to ask questions; instead, they could choose not to do so. Thus, our approach captured end-users' natural curiosity about the driving scenarios they observed.

Our findings from the first study also point to potential limitations of existing XAI mechanisms that focus on explaining outputs of individual AI models. Not only are AI-driven AVs composed of multiple subsystems that require coordination to answer end-user questions, few (if any) such existing XAI mechanisms answer the specific types of questions that our participants asked. Furthermore, even if they support dynamic followup queries (e.g., using interactive ``what-if'' tools~\cite{Wexler2019}), most such tools use visual analytics that end-users might not have expertise with, or that might not be appropriate for interaction inside the vehicle during the ride~\cite{dong2023did, ha2020effects, wiegand2019drive}.

Our second study showed that natural language explanations~\cite{He2025} are a promising direction for answering passengers' questions about AI-driven AVs. Our results showed that static natural language explanations that focus on raising passengers situational awareness can help them understand the driving scenarios they find themselves in. This is particularly important because passengers might not know what questions to ask to comprehensively explore different aspects of each driving scenario. Followup questions to static explanations can help further clarify driving scenarios. Yet,  the key value that conversational XAI interfaces provide is the ability to answer broader questions that help passengers recognize that AVs are AI-driven, teach them how AI works and what it can and cannot do~\cite{long2020ai}.

Our results showed that end-users struggled with technical aspect of AI-driven AV, including their implementation, decision-making steps of different subsystems, and the role of humans in creation, deployment, and running of AVs. Our natural language explanations could not help them with raising such AI literacy competencies either.
Although understanding technical aspects system's decisions could help some end-users to develop a nuanced appreciation for its capabilities and limitations, not every end-user might desire such knowledge. Furthermore, as end-users probe deeper into the inner workings of AI systems, they may encounter complexities that challenge the limits of their technical expertise and understanding. This could potentially lead to misconceptions or concerns about system reliability. Therefore, it is important to strike a balance between transparency and complexity of explanations.

Our results add further evidence that LLMs could help deliver natural language explanations. Not only can they provide a conversational XAI interface that communicates metadata about a model in an end-user interpretable and comprehensible manner~\cite{slack2023explaining}, they can also answer general knowledge questions relevant to AI-driven AV that may not be present in such metadata. However, it is important to note that LLMs can face challenges to provide accurate answers. For example, LLMs can hallucinate wrong or biased responses that end-users struggle to validate~\cite{Prabhudesai2025}.

Although our findings mainly apply to XAI solutions in the autonomous driving domain, they also generalize more broadly to human-centered XAI. Thus, they inform the design of future XAI tools through key design implications:

\begin{itemize}
    \item \textbf{Interactive and Ongoing Explanations}: explanations should be dynamic, allowing users to engage in dialogue-like exchanges with the AI. Allowing follow-up questions is integral, as this could foster not only task and model comprehension, but also appropriate trust.

    \item \textbf{Multi-Model and Subsystem Explanations}: XAI tools should provide holistic explanations of complex, multi-model AI systems across different subsystems they are composed of.

    \item \textbf{Task-Specific Explanations}: Focusing on explaining the task that the AI is trying to perform, instead of simply using explanations to justify AI decisions, can significantly aid end-user understanding of decision-making within AI systems.

    \item \textbf{Support for Persistent Curiosity}: XAI tools should satisfy end-users' continuous quest for learning, helping them develop new and amplify existing AI literacy competencies.

    \item \textbf{Improving the XAI Tool Knowledge Base}: XAI tools should be enhanced with comprehensive knowledge base that goes beyond inputs and outputs to an AI model, and include broader knowledge about the socio-technical context in which the AI operates.
\end{itemize}

Designing future XAI tools following our implications has the potential to democratize explanations for a broader set of end-users of AI-driven AVs. Such tools have the potential to educate the public about this quickly emerging technology in a way that can foster appropriate trust and broader adoption.

\section{Conclusion and Future Work}

In this paper, we explored the AI explanation needs of end-users, specifically in the domain of AI-driven AVs. We conducted two user studies to examine different to identify different ways in which end-users seek answers about driving scenarios and how those answer help them understand those scenarios.
The first study highlighted the need for comprehensive explanations that help end-users build situational awareness, but also satisfy their curiosity about AI knowledge beyond the immediate task, contest undesirable AI behaviors, and seek ways to repair trust.
The second study showed that natural language explanations delivered using conversational XAI interfaces have the potential to improve end-users' understanding of different driving scenarios and raise their overall literacy about AI-driven AVs.

Our work opens up opportunities for future research. One such opportunity is exploring ways to combine LLMs with existing XAI tools. Future research can also explore new ways to merge model-level explanations into system-level ones, creating end-user-friendly, accurate explanations for complex AI systems. Additionally, investigating how our methods can generalize beyond the automotive domain and adapt to different domains in which AI is deployed is crucial for the field's advancement.
Furthermore, extending our question-answering paradigm also presents valuable AI educational opportunities. End-users engaging with AI systems in this manner may not only enhance their understanding of specific tasks but also gain a broader appreciation for AI concepts. This educational component is vital for cultivating a well-informed public in an age where AI increasingly influences daily life.

\begin{acks}
This article is based upon work in part supported by the National Science Foundation (NSF) under Grant No. IIS-2237562. 
\end{acks}

\bibliographystyle{ACM-Reference-Format}
\bibliography{main}

\pagebreak

\appendix

\section{Driving Scenario Metadata}
\autoref{tab:consolidated_summary_unique_values} summarizes the annotated metadata we used in our studies and the possible value for each variable in three types of events: baseline, crash, and near-crash driving events. We only displayed maximum of 5 values to familiarize the reader with the type of scenarios that can be formed from the combination of these variables.
\scriptsize
\begin{longtable}{|p{3cm}|p{3cm}|p{3cm}|p{3cm}|}
    \caption{Consolidated Summary of Unique Values for Key Columns in Baseline, Near Crash, and Crash Data}
    \label{tab:consolidated_summary_unique_values}
    \tabularnewline
    \hline
    \textbf{Column} & \textbf{Baseline Data} & \textbf{Near Crash Data} & \textbf{Crash Data} \\
    
    \endfirsthead
    \hline
    \textbf{Column} & \textbf{Baseline Data} & \textbf{Near Crash Data} & \textbf{Crash Data} \\
    
    \endhead

    \multicolumn{4}{r}{\textit{Continued on the next page...}}
    \endfoot

    \endlastfoot
    \hline
    \textbf{eventNature1} & Not Eventful &  'Conflict with parked vehicle', 'Conflict with pedalcyclist', 'Conflict with animal', 'Conflict with obstacle/object in roadway', 'Conflict with a lead vehicle' &   'Conflict with a lead vehicle', 'Conflict with obstacle/object in roadway', 'Conflict with a following vehicle', 'Single vehicle conflict', 'Conflict with vehicle turning into another vehicle path (same direction)' \\
    \hline
    \textbf{incidentType1} & No Incident & Sideswipe, same direction (left or right), Straight crossing path, Rear-end, struck &   Rear-end, struck, Backing, fixed object, Road departure (end) \\
    
    \textbf{eventSeverity1} & Not Applicable & 'Near-Crash' & 'Crash' \\
    \hline
    \textbf{crashSeverity1} & Not applicable & Not a Crash &  'I - Most Severe', 'IV - Low-risk Tire Strike', 'II - Police-reportable Crash', 'III  - Minor Crash' \\
    \hline
    \textbf{vehicle1EvasiveManeuver1} & Not applicable &  'Braked and steered right', 'Braked (lockup)', 'No reaction', 'Braked (no lockup)', 'Released brakes' &  'Braked and steered right', 'Braked (lockup)', 'Braked and steered left', 'Braked (no lockup)', 'Steered to left' \\
    \hline
    \textbf{vehicle1PostManeuver1} & Not applicable &  'Control maintained', 'Skidded longitudinally', 'Combination of previous' &  'Rotated clockwise', 'Control maintained', 'Skidded longitudinally'\\
    \hline
    \textbf{eventNature2} & None &  'None', 'Conflict with a lead vehicle', 'Conflict with oncoming traffic', 'Conflict with a following vehicle', 'Conflict with vehicle turning into another vehicle path (opposite direction)' & 'None' \\
        \hline
        \textbf{incidentType2} & Not applicable & \raggedright 'Rear-end, struck', 'None', 'Opposite direction (head-on or sideswipe)', 'Road departure (left or right)' & 'None' \\
        \hline
        \textbf{eventSeverity2} & Not applicable & \raggedright 'Not Applicable', 'Crash', 'Near-Crash', 'Non-Subject Conflict', 'Crash-Relevant' & 'Not Applicable' \\
        \hline
        \textbf{crashSeverity2} & Not applicable &  'Not Applicable', 'Not a Crash', 'III  - Minor Crash' & 'Not Applicable' \\
        \hline
        \textbf{vehicle1EvasiveManeuver2} & Not applicable &  'Not Applicable', 'Braked and steered right', 'No reaction', 'Braked (no lockup)', 'Braked and steered left' & 'Not Applicable' \\
        \hline
        \textbf{vehicle1PostManeuver2} & Not applicable & 'Not Applicable', 'Control maintained', 'Combination of previous' & 'Not Applicable' \\
        \hline
        \textbf{airbagDeployment} & No & 'Not Applicable', 'No' & 'No' \\
        \hline
        \textbf{vehicleRollover} & No & 'Not Applicable', 'No' & 'No' \\
        \hline
        \textbf{driverBehavior1} & 'Stop sign violation, "rolling stop"', 'None', 'Driving slowly: below speed limit', 'Other', 'Failed to signal' & 'Aggressive driving, other', 'Signal violation, apparently did not see signal', 'Exceeded safe speed but not speed limit', 'Cutting in, too close behind other vehicle', 'Improper turn, wide left turn' & 'None', 'Improper backing, other', 'Improper turn, cut corner on right', 'Exceeded speed limit', 'Signal violation, apparently did not see signal',  \\
        \hline
        \textbf{driverBehavior2} & 'Other', 'Failed to signal', 'No Additional Driver Behaviors' & 'Improper turn, wide right turn', 'Improper turn, other', 'No Additional Driver Behaviors', 'Apparent general inexperience driving', 'Exceeded speed limit' & 'Improper turn, wide right turn', 'Stop sign violation, apparently did not see stop sign', 'No Additional Driver Behaviors', 'Improper turn, cut corner on right', 'Exceeded speed limit' \\
        \hline
        \textbf{driverBehavior3} & 'No Additional Driver Behaviors' & 'Sudden or improper braking', 'Cutting in, too close in front of other vehicle', 'No Additional Driver Behaviors', 'Passing on right' & 'No Additional Driver Behaviors' \\
        \hline
        \textbf{frontSeatPassengers} & 1, 2 & 1, 2 & 1, 2 \\
        \hline
        \textbf{rearSeatPassengers} & 0, 1, 2 & 0, 1, 2 &0, 1, 2 \\
        \hline
        \textbf{vehicleContributingFactors} & 'None' & 'None' & 'Brake system', 'None' \\
        \hline
        \textbf{visualObstructions} & 'No Obstruction' & 'Parked vehicle', 'Building, billboard, or other roadway infrastructure design features', 'Sunlight', 'Moving vehicle (with or without load)' & 'Moving vehicle (with or without load)', 'No obstruction' \\
        \hline
        \textbf{lighting} & 'Dusk', 'Dawn', 'Darkness, lighted', 'Darkness, not lighted', 'Daylight' & 'Dusk', 'Dawn', 'Darkness, lighted', 'Darkness, not lighted', 'Daylight' & 'Dusk', 'Dawn', 'Darkness, lighted', 'Darkness, not lighted', 'Daylight' \\
        \hline
        \textbf{weather} & 'No Adverse Conditions', 'Mist/Light Rain', 'Raining' & 'No Adverse Conditions', 'Mist/Light Rain', 'Snowing', 'Raining' & 'No Adverse Conditions', 'Snowing', 'Raining', 'Fog', 'Mist/Light Rain' \\
        \hline
        \textbf{surfaceCondition} & 'Dry', 'Wet' & 'Snowy', 'Dry', 'Wet' & 'Snowy', 'Dry', 'Wet' \\
        \hline
        
        \textbf{trafficFlow} & 'Not divided - center 2-way left turn lane', 'Not divided - simple 2-way trafficway', 'Divided (median strip or barrier)', 'One-way traffic' & 'One-way traffic', 'Not divided - center 2-way left turn lane', 'No lanes', 'Divided (median strip or barrier)', 'Not divided - simple 2-way trafficway' & 'No lanes', 'Not divided - simple 2-way trafficway', 'Divided (median strip or barrier)', 'Not divided - center 2-way left turn lane' \\
        
        \hline
        
        \textbf{contigTravelLanes} & 1, 2, 3, 4, 5, 7 & 0, 1, 2, 3, 4, 5 & 0, 1, 2, 3, 4, 5, 6 \\
        \hline
        \textbf{throughTravelLanes} & 0, 1, 2, 3, 4 & 0, 1, 2, 3, 4, 5 & 0, 1, 2, 3, 4 \\
        \hline
        \textbf{infrastructure} & 'None' & 'Weather, visibility', 'None' & 'Weather, visibility', 'Roadway sight distance', 'None' \\
        \hline
        \textbf{trafficDensity} & 'Level-of-service A1: Free flow, no lead traffic', 'Level-of-service A2: Free flow, leading traffic present', 'Level-of-service B: Flow with some restrictions' & 'Level-of-service E: Flow is unstable, vehicles are unable to pass, temporary stoppages, etc.', 'Level-of-service C: Stable flow, maneuverability and speed are more restricted', 'Unknown' & 'Level-of-service C: Stable flow, maneuverability and speed are more restricted',  'Level-of-service D: Unstable flow - temporary restrictions substantially slow driver' \\
        \hline
        \textbf{trafficControl} & 'No traffic control',  'Construction signs/warnings', 'Traffic lanes marked', 'Slow or warning sign, other', 'Traffic signal', 'Stop sign' & 'No traffic control', 'Yield sign', 'Construction signs/warnings', 'Traffic lanes marked', 'Slow or warning sign, other' & 'No traffic control', 'Yield sign', 'Construction signs/warnings', 'Traffic lanes marked', 'Traffic signal' \\
        \hline
        \textbf{relationToJunction} & 'Non-junction', 'Other', 'Intersection-related', 'Driveway, alley access, etc.', 'Intersection', 'Parking lot entrance/exit', 'Interchange area' & 'Non-junction', 'Parking lot, within boundary', 'Intersection-related', 'Intersection', 'Parking lot entrance/exit', 'Interchange area' & 'Non-junction', 'Parking lot, within boundary', 'Intersection-related', 'Driveway, alley access, etc.', 'Intersection', 'Parking lot entrance/exit', 'Interchange area' \\
        \hline
        \textbf{alignment} & 'Curve right', 'Straight', 'Curve left' & 'Curve right', 'Straight', 'Curve left' & 'Curve right', 'Straight', 'Curve left' \\
        \hline
        \textbf{constructionZone} & 'Construction zone-related (occurred in approach or otherwise related to zone)', 'Not construction zone-related', 'Construction Zone (occurred in zone)' & 'Construction zone-related (occurred in approach or otherwise related to zone)', 'Not construction zone-related', 'Construction Zone (occurred in zone)' & 'Not construction zone-related', 'Construction Zone (occurred in zone)' \\
        \hline
        \textbf{numberOfOtherMotorists} & 0 & 0, 1, 2 & 0, 1, 2 \\
        \hline
        \textbf{numberOfObjectsAnimals} & 0 & 0, 1 & 0, 1 \\
        \hline
        \textbf{fault} & 'Not applicable' & 'Not applicable', 'Subject driver', 'Driver 2' & 'Not applicable', 'Subject driver', 'Driver 2'\\
        \hline
        \textbf{intersectionInfluence} & 'Yes, Stop Sign', 'No', 'Yes, Other', 'Yes, Parking lot, Driveway Entrance/Exit', 'Yes, Uncontrolled', 'Yes, Traffic Signal' & 'Yes, Stop Sign', 'No', 'Yes, Other', 'Yes, Parking lot, Driveway Entrance/Exit', 'Yes, Interchange', 'Yes, Uncontrolled', 'Yes, Traffic Signal' & 'Yes, Stop Sign', 'No', 'Yes, Other', 'Yes, Parking lot, Driveway Entrance/Exit', 'Yes, Uncontrolled', 'Yes, Traffic Signal' \\
        \hline
        \textbf{preIncidentManeuver} & 'Going straight, constant speed', 'Decelerating in traffic lane', 'Turning right', 'Maneuvering to avoid a pedestrian/pedalcyclist', 'Turning left', 'Changing lanes' & 'Turning left', 'Stopped in traffic lane', 'Changing lanes', 'Starting in traffic lane', 'Going straight, but with unintentional "drifting" within lane or across lanes' & 'Going straight, constant speed', 'Decelerating in traffic lane', 'Turning right', 'Turning left', 'Starting in traffic lane', 'Changing lanes' \\
        \hline
        \textbf{maneuverJudgment} & 'Unsafe and illegal', 'Safe and legal', 'Safe but illegal', 'Unsafe but legal' & 'Unsafe and illegal', 'Safe and legal', 'Unsafe but legal' & 'Unsafe and illegal', 'Safe and legal', 'Unsafe but legal' \\
        \hline    
\end{longtable}

\normalsize

\section{Appendix: Post Event Questionnaire}
\label{app:post-event-questions}

The driving scenario comprehension quiz contained questions that correspond to expert-annotated metadata for each driving scenario. Order of MCQ choices was randomized. The correctness of each question depended on the metadata of a given driving scenario.

\scriptsize

\begin{enumerate}[label=Q\arabic*)]

    \item What did the autonomous vehicle do in this driving event?

        \begin{itemize}
            \item[A)] Crash
            \item[B)] Near-crash
            \item[C)] I can not identify
        \end{itemize}
        
    \item Did the autonomous vehicle roll over?

        \begin{itemize}
            \item[A)] Yes
            \item[B)] No
           
        \end{itemize}
    \item Did the autonomous vehicle fail to use its break system?

        \begin{itemize}
            \item[A)] Yes
            \item[B)] No
          
        \end{itemize}
    \item Was there anything that obstructs the given vehicle’s view?
 
        \begin{itemize}
            \item[A)] Rain, snow, fog, smoke, sand, dust
            \item[B)] Building, billboard, or other roadway infrastructure design features
            \item[C)] Moving vehicle (with or without load)
            \item[D)] No obstruction interfered with the autonomous vehicle ability to drive
        \end{itemize}
    \item What lighting condition the autonomous vehicle was driving in?

        \begin{itemize}
            \item[A)] Daylight
            \item[B)]Darkness, lighted
            \item[C)] Dawn
            \item[D)] Dusk
        \end{itemize}
    \item What was the autonomous vehicle doing that might have led to the incident? Select the best answer.
        \begin{itemize}
            \item[A)]Passing or overtaking another vehicle
            \item[B)] Making U-turn
            \item[C)] Backing up (other than for parking purposes)
            \item[D)] Stopped in traffic lane
        \end{itemize}
    \item How do you describe the nature of what the autonomous vehicle did?
 
        \begin{itemize}
            \item[A)] Conflict with vehicle moving across another vehicle path (through intersection)
            \item[B)] The autonomous vehicle itself was the only one involved
            \item[C)] Conflict with vehicle turning into another vehicle path (opposite direction)
            \item[D)] Conflict with obstacle/object in roadway
        \end{itemize}
    \item How severe was the incident involving the autonomous vehicle?

        \begin{itemize}
            \item[A)] Police-reportable Crash
            \item[B)] Not a Crash
            \item[C)] Minor Crash
            \item[D)] Most Severe
        \end{itemize}
    \item Which one of the following evasive behaviors did the autonomous vehicle do in the given driving event?

        \begin{itemize}
            \item[A)] Braked (no lockup)
            \item[B)] Braked and steered right
            \item[C)] Braked (lockup)
            \item[D)] Braked and steered left
        \end{itemize}
    
    \item What did the autonomous vehicle do after the incident?

        \begin{itemize}
            \item[A)] Rotated clockwise
            \item[B)] Skidded longitudinally
            \item[C)] Control maintained
            \item[D)] Combination of previous
        \end{itemize}
    \item How was the weather condition in the given driving event?
   \begin{itemize}
        \item[A)] Raining
        \item[B)] Fog
        \item[C)] Snowing
        \item[D)] No Adverse Conditions
    \end{itemize}
    \item Which one of the following weather conditions present in this scenario might have interfered with the autonomous vehicle’s driving?

 \begin{itemize}
        \item[A)] Dirt road
        \item[B)] Dry
        \item[C)] Snowy
        \item[D)] Wet
    \end{itemize}

 \item What type of road was the autonomous vehicle driving on?
    \begin{itemize}
        \item[A)] One-way traffic
        \item[B)] Not divided - simple 2-way traffic-way
        \item[C)] Not divided - center 2-way left turn lane
        \item[D)] Divided (median strip or barrier)
    \end{itemize}
    \item Which of the following traffic densities might have influenced the autonomous vehicle’s driving?
    \begin{itemize}
        \item[A)] Flow is unstable, vehicles are unable to pass, temporary stoppages, etc.
        \item[B)] Flow with some restrictions
        \item[C)] Free flow, no lead traffic
        \item[D)] Free flow, leading traffic present
    \end{itemize}
    \item Which one of the following traffic control signs should the autonomous vehicle be able to detect in this driving event?
    \begin{itemize}
        \item[A)] No traffic control
        \item[B)] School zone related sign
        \item[C)] Slow or warning sign, other
        \item[D)] Yield sign
    \end{itemize}
    \item How was the alignment of the autonomous vehicle with the road while driving?
    \begin{itemize}
        \item[A)] Curve left
        \item[B)] Curve right
        \item[C)] Straight
        \item[D)] None of the above
    \end{itemize}
    \item Did the construction zone contribute to autonomous vehicles’ behavior?
    \begin{itemize}
        \item[A)] Construction Zone (occurred in zone)
        \item[B)] Construction zone-related (occurred in approach or otherwise related to zone)
        \item[C)] Not construction zone-related
        \item[D)] I can not determine
    \end{itemize}
    \item How many other motorists did contribute to this driving event?

    \begin{itemize}
        \item[A)] 0
        \item[B)] 1
        \item[C)] 2
        \item[D)] 3
    \end{itemize}
    \item Were there any other objects or animals on the road that were involved in what happened in the given driving event?

    \begin{itemize}
        \item[A)] Yes
        \item[B)] No
       
    \end{itemize}
    \item Who was at fault in the given driving event?
    \begin{itemize}
        \item[A)] Not applicable
        \item[B)] Unable to determine
        \item[C)] Driver of the other car
        \item[D)] The autonomous vehicle
    \end{itemize}

    \item How was the autonomous vehicle's maneuver judgment?
    \begin{itemize}
        \item[A)] Unsafe but legal
        \item[B)] Unsafe and illegal
        \item[C)] Safe and legal
        \item[D)] Safe but illegal
    \end{itemize}
 
\end{enumerate}
\normalsize

\section{AI Literacy Questionnaire}
\label{app:ai-literacy-event-questions}

The AI literacy assessment quiz contained questions that correspond to each of the 17 AI literacy competencies grouped by category and competency from~\cite{long2020ai}. Order of MCQ choices was randomized. Each multiple choice question has the correct answer in bold. 

\textbf{What is AI?}

\scriptsize

\begin{enumerate}[label=C\arabic*:]
    \item Recognizing AI.
    \begin{enumerate}[label=Q\arabic*)]
        \item Autonomous Vehicles use AI to operate. True or false?
        \begin{enumerate}[label=\Alph*)]
            \item \textbf{True.}
            \item False.
        \end{enumerate}
    \end{enumerate}
    \item Understanding Intelligence.
    \begin{enumerate}[label=Q\arabic*)]\addtocounter{enumii}{1}
        \item What kind of intelligence do autonomous vehicles exhibit? Select the best answer.
        \begin{enumerate}[label=\Alph*)]
            \item \textbf{Human Intelligence.}
            \item Machine (or artificial) intelligence.
            \item Animal intelligence.
            \item None of the above.
        \end{enumerate}
    \end{enumerate}
    \item Interdisciplinary.
    \begin{enumerate}[label=Q\arabic*)]\addtocounter{enumii}{2}
        \item Which of the following are different approaches or perspectives in developing ``intelligent machines'' for autonomous cars? Select all that apply.
        \begin{enumerate}[label=\Alph*)]
            \item \textbf{Rule-Based Systems.}
            \item \textbf{Neural Networks and Machine Learning.}
            \item Manual Control and Direct Input.
            \item Genetic Modification and Cloning.
        \end{enumerate}
    \end{enumerate}

    \item General vs. Narrow.
    \begin{enumerate}[label=Q\arabic*)]\addtocounter{enumii}{3}
        \item  What best characterizes autonomous cars? Select the best answer.
        \begin{enumerate}[label=\Alph*)]
            \item \textbf{They are characterized by specialized skills and capabilities limited to specific tasks or domains.}
            \item They are characterized by a broad range of cognitive abilities and adaptability in various tasks and contexts.
          \item They are characterized by the ability to navigate and interact effectively in social situations.
          \item None of the above.
        \end{enumerate}
    \end{enumerate}
\end{enumerate}
\normalsize

\textbf{What Can AI Do?}

\scriptsize
\begin{enumerate}[label=C\arabic*:]\addtocounter{enumi}{4}
    \item AI's Strengths \& Weaknesses.
    \begin{enumerate}[label=Q\arabic*)]\addtocounter{enumii}{4}
        \item Which of the following is a problem that an autonomous car excels at solving? Select the best answer.
        \begin{enumerate}[label=\Alph*)]
            \item \textbf{Parking in a tight spot.}
            \item Driving in heavy traffic.
            \item Driving in a snowstorm.
            \item Talking to the passengers in the car.
        \end{enumerate}

        \item Which of the following is a problem that an autonomous car may have difficulty solving? Select the best answer.
       \begin{enumerate}[label=\Alph*)]
          \item \textbf{Keeping distance from the car in front of them on the road.}
          \item Staying in their lane on the road.
          \item Detecting objects in heavy rain.
          \item None of the above.
      \end{enumerate}
    \end{enumerate}

    \item Imagine Future AI.
    \begin{enumerate}[label=Q\arabic*)]\addtocounter{enumii}{6}
        \item Transportation companies invest in autonomous vehicles in part because of the promise that they could replace human professional drivers (e.g., truck drivers, taxi drivers, etc.) in the future. True or false?
        \begin{enumerate}[label=\Alph*)]
            \item \textbf{True.}
            \item False.
        \end{enumerate}

        \item AI technologies for autonomous driving will be ready for widespread adoption and presence on the roads in the next 5 years. True or false?
        \begin{enumerate}[label=\Alph*)]
            \item True.
            \item \textbf{False.}
        \end{enumerate}
    \end{enumerate}
\end{enumerate}
\normalsize

\textbf{How Does AI Work?}

\scriptsize
\begin{enumerate}[label=C\arabic*:]\addtocounter{enumi}{6}
    \item Representation.
    \begin{enumerate}[label=Q\arabic*)]\addtocounter{enumii}{8}
        \item How does the image of a stop sign is represented to an autonomous vehicle?
        \begin{enumerate}[label=\Alph*)]
            \item An array of numbers that represent the intensity of three colors (red, blue, green)
            \item A set of information about the color, shape, and text of the sign.
            
        \end{enumerate}
        \item How does the current autonomous vehicle get input from the environment? Select the best answer.
        \begin{enumerate}[label=\Alph*)]
          \item \textbf{Data from sensors.}
          \item Data entered by the passenger of the autonomous vehicle.
          \item By analyzing satellite imagery to understand the surrounding environment.
          \item By receiving direct input from other autonomous vehicles in the vicinity.
      \end{enumerate}
    \end{enumerate}

    \item Decision-Making.
    \begin{enumerate}[label=Q\arabic*)]\addtocounter{enumii}{10}
        \item Autonomous vehicles adjust their driving relative to weather conditions. True or false?
        \begin{enumerate}[label=\Alph*)]
            \item \textbf{True.}
            \item False.
        \end{enumerate}

        \item Autonomous vehicles can recognize the other vehicle stopped in front of them. True or false?
        \begin{enumerate}[label=\Alph*)]
            \item \textbf{True.}
            \item False.
        \end{enumerate}

        \item Autonomous vehicles can detect stop signs. True or false?
        \begin{enumerate}[label=\Alph*)]
            \item \textbf{True.}
            \item False.
        \end{enumerate}
        
        \item How do autonomous vehicles make decisions? Select the best answer.
        \begin{enumerate}[label=\Alph*)]
            \item \textbf{By using complex algorithms and machine learning techniques to analyze sensor data and make decisions based on patterns and correlations.}
          \item By directly translating human language into computer-readable formats to enable comprehension.
          \item By relying solely on pre-programmed rules and logic to interpret and understand the world.
          \item By using a network of interconnected supercomputers to simulate and model the entire world in real time.
        \end{enumerate}
    \end{enumerate}

    \item Machine Learning Steps.
    \begin{enumerate}[label=Q\arabic*)]\addtocounter{enumii}{14}
        \item Which of the following is not a step required to enable autonomous vehicles to drive? Select the best answer.
        \begin{enumerate}[label=\Alph*)]
            \item \textbf{Program how the vehicle should behave in each unique situation.}
            \item Performing extensive road testing to collect real-world driving data.
            \item Assigning labels to the vehicle-specific and driving-specific data by experts.
            \item Train AI models for subsystems of autonomous vehicles and test them.
          
        \end{enumerate}
    \end{enumerate}

    \item Human Role in AI.
    \begin{enumerate}[label=Q\arabic*)]\addtocounter{enumii}{15}
        \item What is the primary role of humans riding in fully autonomous vehicles? Select the best answer.
        \begin{enumerate}[label=\Alph*)]
            \item \textbf{Taking control of the vehicle in emergency situations.}
            \item Communicating with the autonomous vehicle.
            \item Guide the autonomous vehicle through the construction zone.
            \item Steering the wheel to direct the vehicle.
        \end{enumerate}

        \item What is the primary role of humans in ensuring autonomous vehicles operate correctly? Select all that apply.
        \begin{enumerate}[label=\Alph*)]
            \item \textbf{Designing and training the AI models used in autonomous vehicles.}
          \item \textbf{Monitoring and supervising autonomous vehicles during their test operation.}
          \item Supervising autonomous vehicles at all times.
          \item Physically operating the autonomous vehicles during their operation.
        \end{enumerate}
    \end{enumerate}

    \item Data Literacy.
    \begin{enumerate}[label=Q\arabic*)]\addtocounter{enumii}{17}
        \item How does the AV detect objects in the surrounding area? Select all that apply.
        \begin{enumerate}[label=\Alph*)]
            \item \textbf{Using laser light reflected off of surrounding objects.}
          \item Using magnetic waves reflected off of surrounding objects.
          \item Using sounds emitted from the surrounding objects.
          \item Using input from the passengers.
        \end{enumerate}

        \item The decision making subsystem in autonomous vehicles have the ability to learn from the data. True or false?
        \begin{enumerate}[label=\Alph*)]
            \item \textbf{True.}
            \item False.
        \end{enumerate}

        \item What data do autonomous vehicles (AV) learn from to detect and identify objects in their surroundings, including pedestrians? Select the best answer.
        \begin{enumerate}[label=\Alph*)]
            \item \textbf{Examples of similar objects that were previously collected using their sensors and labeled and categorized by humans.}
          \item Passengers’ description of different objects in their surroundings.
          \item Existing computer program rules that AV programmers specified for each object.
          \item Images of the objects available on the Internet.
        \end{enumerate}
    \end{enumerate}

    \item Learning from Data.
    \begin{enumerate}[label=Q\arabic*)]\addtocounter{enumii}{20}
        \item The autonomous vehicles learn how to drive from existing driving data. True or false?
        \begin{enumerate}[label=\Alph*)]
            \item \textbf{True.}
            \item False.
        \end{enumerate}
    \end{enumerate}

    \item Critically Interpreting Data.
    \begin{enumerate}[label=Q\arabic*)]\addtocounter{enumii}{21}
        \item Driving data collected in the US can be used to train autonomous vehicles elsewhere in the world. True or false?
        \begin{enumerate}[label=\Alph*)]
            \item True.
            \item \textbf{False.}
        \end{enumerate}
    \end{enumerate}

    \item Action \& Reaction.
    \begin{enumerate}[label=Q\arabic*)]\addtocounter{enumii}{22}
        \item Which of the following statements is true about the ability of autonomous vehicles systems to act on the world physically? Select the best answer. Hint: Example of acting on the world: A robot goes from point A to point B. Example of reacting to the world: A robot jumps to avoid an obstacle.
        \begin{enumerate}[label=\Alph*)]
            \item \textbf{Autonomous vehicles can both react to the world and act on the world in a planned and deliberate manner.}
            \item Autonomous vehicles can only react to the world.
            \item Autonomous vehicles can only act on the world in a planned and deliberate manner.
            \item Autonomous vehicles cannot act on the world at all.
        \end{enumerate}
    \end{enumerate}

    \item Sensors.
    \begin{enumerate}[label=Q\arabic*)]\addtocounter{enumii}{23}
        \item Which of the following are used by autonomous vehicles to perceive the world around them? Select the best answer.
        \begin{enumerate}[label=\Alph*)]
            \item Sensors that use light to create a picture of the environment.
          \item Sensors that use radio waves to detect objects.
          \item Sensors that use lasers to create a 3D map of the environment.
          \item \textbf{All of the above.}
        \end{enumerate}
    \end{enumerate}
\end{enumerate}
\normalsize

\textbf{How Should AI Be Used?}

\scriptsize
\begin{enumerate}[label=C\arabic*:]\addtocounter{enumi}{15}
    \item Ethics.
    \begin{enumerate}[label=Q\arabic*)]\addtocounter{enumii}{24}
        \item Which of the following is a potential ethical issue surrounding AI? Select all that apply.
        \begin{enumerate}[label=\Alph*)]
            \item \textbf{AI systems can collect and store large amounts of data about individuals, which could be used to track their movements, monitor their activities, or even predict their future behavior.}
          \item \textbf{AI systems are capable of automating many tasks that are currently performed by humans, which could lead to job losses and economic disruption.}
          \item It is always clear how AVs are making their decisions and who is accountable for those decisions.
          \item AI systems are trained on data that is collected from the real world, which ensures that AVs are not biased by developer decisions.
        \end{enumerate}
    \end{enumerate}
\end{enumerate}
\normalsize

\textbf{How Do People Perceive AI?}

\scriptsize
\begin{enumerate}[label=C\arabic*:]\addtocounter{enumi}{16}
    \item Programmability.
    \begin{enumerate}[label=Q\arabic*)]\addtocounter{enumii}{25}
        \item Autonomous vehicles are programmable. True or false?
        \begin{enumerate}[label=\Alph*)]
            \item \textbf{True.}
            \item False.           
        \end{enumerate}
    \end{enumerate}

\end{enumerate}

\normalsize

\end{document}